\documentclass{jpp}
\usepackage{graphicx}
\usepackage{hyperref}
\usepackage{tikz}

\usepackage[utf8]{inputenc}
\usepackage[T1]{fontenc}
\usepackage{amsmath}
\usepackage{amssymb}
\usepackage{caption}
\usepackage{subcaption}
\usepackage{booktabs}
\usepackage{float}

\hypersetup{
    colorlinks=true,
    linkcolor=blue,
    urlcolor=blue,
    citecolor=blue}

    \shorttitle{Double-Adiabatic Equations of State for Relativistic Plasmas}
\shortauthor{A. Wierzchucka et al.}

\title{Double-Adiabatic Equations of State for Relativistic Plasmas}

\author{A. Wierzchucka\aff{1,2}
  \corresp{\email{agnieszka.wierzchucka@merton.ox.ac.uk}},
   P. J. Bilbao\aff{1,3}, 
   A. G. R. Thomas\aff{4}, \\
   D. A. Uzdensky\aff{1}
   \and  A. A. Schekochihin\aff{1,2}}

\affiliation{\aff{1}Rudolf Peierls Centre for Theoretical Physics, University of Oxford, Oxford, OX1 3PU, UK
\aff{2}Merton College, Oxford OX1 4JD, UK
\aff{3}Lady Margaret Hall, Oxford, OX2 6QA, UK
\aff{4}Gérard Mourou Center for Ultrafast Optical Science, University of Michigan, 2200 Bonisteel Boulevard, Ann Arbor, 48109, Michigan, USA}

\newcommand{\B}[1]{{\mathbf{#1}}}

\begin{document}

\maketitle

\begin{abstract}
The adiabatic equation of state $P \propto n^{\Gamma}$ describes the pressure evolution of highly collisional, isotropic plasmas in terms of their density, providing a possible closure of the fluid moment hierarchy in the absence of heat fluxes and dissipation. An analogous closure exists for collisionless, magnetised plasmas, whose pressure tensor is anisotropic with respect to the magnetic field, and the closure is therefore double adiabatic, prescribing the evolution of the parallel and perpendicular pressures in terms of the magnetic-field strength and density. Here, we present a general first-principles formalism to derive adiabatic laws using symmetries and phase-space conservation properties of the distribution function. With this theory we recover the adiabatic equation of state $P \propto n^{\Gamma}$ for isotropic plasmas and the double-adiabatic equations of state for collisionless, magnetised plasmas. We extend the latter to the relativistic regime, finding that their exact functional form depends on the pressure anisotropy and is not a simple power law. Our double-adiabatic equations of state describe simple geometries, like magnetic mirrors or compressed homogeneous plasmas, as well as complex high-energy astrophysical processes, such as the evolution of plasmoid structures formed during magnetic reconnection. 

\end{abstract}

\section{Introduction}\label{sec1}
In magnetised, collisionless plasmas, constituent particles gyrate rapidly around magnetic-field lines while experiencing infrequent Coulomb collisions. Consequently, the particle gyration frequency sets the shortest timescale in the system, whereas the Coulomb collision time is the largest. Studies of collisionless and weakly collisional magnetised plasmas in the non-relativistic regime have provided significant insight into the dynamics governing space and astrophysical environments such as the solar wind \citep{Hellinger_etal-2006, Bale_etal-2009}, the Earth's magnetosphere \citep{Oieroset_etal-2001, Burch_etal-2016}, and the intracluster medium \citep{Schekochihin_Cowley-2006, Kunz_etal-2022}. Collisionless plasmas are also abundant in high-energy relativistic astrophysical systems, like pulsar-wind nebulae \citep{Kargaltsev_etal-2015}, pulsar magnetospheres \citep{Cerutti_Beloborodov-2017, Philippov_Kramer-2022}, accretion flows onto black holes \citep{Rees_etal-1982, Narayan_Yi-1995, Quataert-2003, Yuan_Narayan-2014}, and jets from Active Galactic Nuclei (e.g., blazars) \citep{Begelman_etal-1984, Blandford_etal-2019}. Many of these sources exhibit pronounced non-thermal emission spectra, providing direct observational evidence that the associated plasmas are far from Maxwellian equilibrium. Such departure from thermal equilibrium means that studying collisionless magnetised plasmas is challenging and requires a kinetic approach. 

Rapid gyromotion renders the particle distribution function azimuthally symmetric in velocity space about the magnetic field, a property known as gyrotropy. The pressure tensor in such a plasma is diagonal. When the magnetic field varies on timescales much longer than the gyration period, the first adiabatic invariant, $\mu$, which is the angular momentum of a particle on a Larmor orbit, is conserved \citep{Gardner-1959}, leading to a non-identical evolution of the perpendicular and parallel pressures, $P_\perp$~and~$P_\parallel$. In the non-relativistic limit, this evolution is described by the Chew--Goldberger--Low (CGL) equations \citep{Chew_etal-1997}, which express the Lagrangian evolution of $P_\perp$ and $P_\parallel$ in terms of the particle number density, $n$, and magnetic-field strength, $B$, as well as heat fluxes. When the latter can be neglected, the CGL equations are known as the double-adiabatic approximation and provide a practical fluid closure of the system: in a moving fluid element, $P_\perp$ and $P_\parallel$ are linked to the density and magnetic-field strength by instantaneous adiabatic equations of state (EoSs), $P_\perp \propto nB$ and $P_\parallel \propto n^3/B^2$. These relations are the anisotropic generalisation of the familiar adiabatic EoS $P \propto n^{\Gamma}$ of a highly collisional, isotropic plasma, where the adiabatic index $\Gamma$ ranges from $5/3$ in the non-relativistic regime to $4/3$ in the ultra-relativistic limit \citep{Synge-1957}.

When heat fluxes can be neglected, the CGL equations are sometimes referred to as ‘double-adiabatic’ because they encompass the conservation of two adiabatic invariants, the first being the aforementioned~$\mu$. The second adiabatic invariant, $J$, is associated with the trapping and bouncing of particles in magnetic mirrors and remains conserved under the condition that this bouncing is much faster than the fluid motions. Such a requirement is much more restrictive than that of rapid gyration, and so the conservation of $J$ is less robust than the conservation of~$\mu$. The conservation of these two quantities can be shown through the use of action–angle variables \citep{Gardner-1959, Kruskal-1962} or directly by expanding the particle equations of motion in small gyration periods, an approximation known as drift kinetics \citep{Northrop_Teller-1960, Vandervoort-1960, Northrop_1963}.

The double-adiabatic equations give insight into the various processes governing the dynamics of collisionless plasmas. As they predict different evolution of $P_\perp$ and $P_\parallel$, local changes in the density or magnetic field, e.g., due to shearing or compressing flows, can give rise to local pressure anisotropies \citep{Hellinger_Travnicek-2008, Kunz_etal-2014, Sironi_Narayan-2015}. In high-$\beta$ plasmas, where $\beta$ is the ratio of the thermal to magnetic pressure, pressure anisotropy can lead to the excitation of various kinetic instabilities (see \citealp{Bott_etal-2024} and references therein). When the anisotropy, $\Delta \equiv P_\perp/P_\parallel - 1$, is positive, the mirror \citep{Southwood_Kivelson-1993} and ion-cyclotron \citep{Hasegawa-1969} instabilities emerge if $\Delta \gtrsim 1/\beta$. In the opposite case, if $\Delta \lesssim -2/\beta$, the microscale dynamics of the plasma becomes dominated by the firehose instability \citep{Chandrasekhar_etal-1958}. After a short period of exponential growth, both the mirror and firehose instabilities saturate and, albeit through different mechanisms, pin the pressure anisotropy to their respective thresholds \citep{Melville_etal-2016, Bott_etal-2025}. As a result, the pressure becomes nearly isotropic and evolves with an adiabatic index $\Gamma = 5/3$ in the high-$\beta$ non-relativistic regime. 

As the distribution function is gyrotropic to lowest order in the gyration period, the Vlasov equation can be averaged over the gyromotion to obtain the so-called drift-kinetic equation. This constitutes a starting point for studying the kinetic instabilities excited in collisionless plasmas, although finite-gyroradius effects must be included to recover the fastest-growing modes \citep{Yoon_etal-1993, Hellinger-2007, Bott_etal-2024}. 

The CGL equations are derived by taking perpendicular and parallel pressure moments of the drift-kinetic equation (see e.g., \citealt{Schekochihin_etal-2010} for a derivation). This procedure can be extended to relativistic collisionless plasmas \citep{Gedalin_Oiberman-1995, Hegade_Stone-2026, Wierzchucka_etal-2026}. Unlike in the non-relativistic regime, the relativistic CGL equations do not provide an evolution equation for the two pressures. Instead, they evolve related quantities that we refer to below as the modified pressures. However, as it is the pressures, and not the modified pressures, that are needed in the fluid momentum equation, this means that, relativistically, the generalised CGL equations cannot provide a sufficient fluid closure of the system. Describing the evolution of $P_\perp$ and $P_\parallel$ in a relativistic, collisionless, magnetised plasma is somewhat non-trivial. The main reason for this is that, in the relativistic case, the momentum-space integration in the definition of $P_\perp$ and $P_\parallel$ involves, in addition to the corresponding momentum components, the particle Lorentz factor $\gamma$, which is a non-linear function of the perpendicular and parallel momenta. As a result, $P_\perp$ ($P_\parallel$) depends not only on the perpendicular (parallel) component of the particle momentum but also on the total momentum (through the Lorentz factor). In other words, each pressure component depends on the full momentum distribution rather than solely on its corresponding momentum projection, as in the non-relativistic case. Therefore, their evolution in the double-adiabatic limit, i.e., in the absence of heat fluxes, turns out to depend not only on the density and magnetic-field strength but also on the pressure anisotropy. 

In this paper, we derive the double-adiabatic equations satisfied by $P_\perp$ and $P_\parallel$ in relativistic collisionless plasmas, with a non-relativistic bulk flow. We do this by abandoning the standard moment-based derivation of the evolution equations and instead formulating a theory of adiabatic invariance focused on the system's symmetries. Through the use of phase-space volume conservation properties, we show that imposing certain symmetries leads to a self-similar evolution of the distribution function. Calculating the pressure moments of the evolved distribution then yields the adiabatic EoSs, whose form, in general, depends on the distribution at some initial (reference) time. For example, one can easily show that a distribution function isotropic in peculiar momentum satisfies $P \propto n^{5/3}$ and $P \propto n^{4/3}$ in the non-relativistic and ultra-relativistic regimes, respectively. If the distribution function is gyrotropic and has parity symmetry along the magnetic-field direction in the peculiar-momentum coordinate, we find by the same procedure that the double-adiabatic equations, $P_\perp \propto nB$ and $P_\parallel \propto n^3/B^2$, hold in the non-relativistic limit.

Using this symmetry-based approach, we extend the double-adiabatic equations to the relativistic regime. We study these equations for the case of an ultra-relativistically hot plasma with an isotropic reference distribution. We find that the exact form of the double-adiabatic equations depends on the pressure anisotropy and derive the asymptotic theory for three main limiting cases. When $P_\perp \gg P_\parallel$ ($\Delta \gg 1)$, we find $P_\perp \propto n B^{1/2}$ and $P_\parallel \propto n^3/B^{5/2}$, whereas when $P_\parallel \gg P_\perp$ ($\Delta \approx -1$), $P_\perp \propto B^2 \ln(n'^2/B'^3)$ and $P_\parallel \propto n^2/B$. Here, $B'$ and $n'$ are respectively the relative changes in the magnetic-field strength and density. These evolution equations agree with the initial discussion of the relativistic CGL equations in \cite{Gedalin_Oiberman-1995}, up to the logarithmic correction in the second $P_\perp$ expression. In contrast, close to isotropy, i.e., when $|\Delta| \ll 1$, we find $P_\perp \propto (nB)^{4/5}$ and $P_\parallel \propto (n^{3}/B^{2})^{4/5}$. Our results differ drastically from the non-relativistic theory \citep{Chew_etal-1997}, in which the same double-adiabatic relations hold for all anisotropies. We confirm our predictions using two-dimensional particle-in-cell (PIC) simulations incorporating a large-scale compressive flow.

In situations where heat fluxes are negligible, these double-adiabatic equations can be the EoSs for relativistic collisionless plasmas. Such EoSs are crucial for modelling large-scale high-energy astrophysical systems by means of a fluid theory, e.g., to determine regions where kinetic processes like the firehose and mirror instabilities are excited in high-$\beta$ plasmas. Their applicability to the relativistic regime also means that our EoSs can be coupled to other high-energy processes, such as synchrotron cooling, providing a pathway to modelling phenomena like the recently discovered synchrotron firehose instability~\citep{Zhdankin_etal-2023}. Double-adiabatic EoSs may additionally have an important effect on the large-scale magnetohydrodynamic-equilibrium structure of astrophysical magnetospheres and flows \citep{Beskin_Kuznetsova-2000}.

\section{Adiabatic Invariance and Symmetries}
We consider a particle distribution function $f(\mathbf{r}, \mathbf{p}, t)$, where $\mathbf{r}$ and $\mathbf{p}$ denote the position and momentum variables, and $t$ denotes time. The distribution function is the density of a phase-space fluid filling the six-dimensional $(\mathbf{r}, \mathbf{p})$ space, moving with velocity 
\begin{equation}
    \label{eq: phase-space velocity}
    \dot{\mathbf{r}} = \mathbf{v}, \quad \dot{\mathbf{p}} = \mathbf{F}.
\end{equation}
The velocity coordinate $\mathbf{v}$ is related to the momentum through $\mathbf{p} = \gamma m \mathbf{v}$, where ${\gamma = \sqrt{1+(p/mc)^2}}$ is the Lorentz factor, and $m$ is the particle mass. The force $\mathbf{F}$ depends on the particular physical system that the distribution function describes.  In this paper, we focus on Hamiltonian systems, in which case the phase-space fluid has the additional property of being incompressible.

To describe how $f$ evolves in time, we adopt a Lagrangian formalism \citep{Newcomb-1961}. We describe the position of an infinitesimal fluid element in phase space using not its current coordinates $(\mathbf{r}, \mathbf{p})$, but instead its position at $t=0$, $(\mathbf{r}_0, \mathbf{p}_0)$. In this approach, the conservation of probability is
\begin{equation}
    \label{eq: phase-space cont}
    d\mathbf{r}d\mathbf{p} \ f(t, \mathbf{r}, \mathbf{p}) = d\mathbf{r}_0d\mathbf{p}_0 \ f_0(\mathbf{r}_0, \mathbf{p}_0),
\end{equation}
and the incompressibility of the phase-space fluid, also referred to as the conservation of phase-space volume, is written as 
\begin{equation}
    \label{eq: full phase space J}
    \left| \frac{\partial (\mathbf{r}, \mathbf{p})}{\partial (\mathbf{r}_0, \mathbf{p}_0)}\right| = 1.
\end{equation}
One can combine \eqref{eq: phase-space cont} and \eqref{eq: full phase space J} to find Liouville's theorem
\begin{equation}
\label{eq: Liouville}
    f(t, \mathbf{r}, \mathbf{p}) = f_0(\mathbf{r}_0, \mathbf{p}_0),
\end{equation}
where $f_0$ is the distribution function at $t=0$. 

Two important moments of $f$ are the density and the \cite{Eckart-1940} bulk velocity, defined, respectively, by
\begin{align}
    \label{eq: bulk def}
    n \equiv \int d \mathbf{p} \ f, \quad
    n\mathbf{u} \equiv \int d\mathbf{p} \ \mathbf{v}f.
\end{align}
In this work, for simplicity, we restrict our analysis to the non-relativistic limit as far as the bulk velocity of the plasma is concerned, $|\mathbf{u}| \ll c$. Nevertheless, our findings regarding the adiabatic and double-adiabatic EoSs remain valid in the fully relativistic regime, as discussed in Section~\ref{sec: Discussion}. We label quantities in the comoving frame with a tilde, referring to them as ‘peculiar’. To first order in $|\mathbf{u}|/c$, the Lorentz transformation of the momentum to the peculiar momentum is
\begin{equation}
    \label{eq: Isotropic momentum split}
    \mathbf{p} = \tilde{\gamma} m \mathbf{u} + \tilde{\mathbf{p}},
\end{equation}
and the corresponding transformation of the Lorentz factor is
\begin{equation}
    \label{eq: Isotropic gamma split}
    \gamma = \frac{\mathbf{u} \cdot \tilde{\B{p}}}{m c^2} + \tilde{\gamma}.
\end{equation}
The Jacobian of this change of variables is, again to first order in $|\mathbf{u}|/c$,
\begin{equation}
    \label{eq: peculiar J}
    \left| \frac{\partial\mathbf{p}}{\partial \tilde{\mathbf{p}}} \right| = 1 +  \frac{\mathbf{u} \cdot \tilde{\B{v}}}{ c^2},
\end{equation}
where $\tilde{\B{v}} = \tilde{\B{p}}/\tilde{\gamma}m$. In the fully non-relativistic regime, $|\tilde{\mathbf{p}}| \ll m c$, \eqref{eq: Isotropic momentum split} reduces to $\mathbf{p} = m\mathbf{u} + \tilde{\mathbf{p}}$, as expected. Note that, while by definition of the bulk velocity \eqref{eq: bulk def}, $\int d\tilde{\mathbf{p}} \ \tilde{\mathbf{v}}f = 0$, in general $\int d\tilde{\mathbf{p}} \ \tilde{\mathbf{p}}f \neq 0$, reflecting the differences between the Landau \citep{Landau_Lifshitz-1987} and Eckart bulk velocities.

\subsection{Isotropy}
\label{sec: isotropy}
For illustration, we first consider a distribution function that evolves in such a way as to remain isotropic in the frame moving with $\mathbf{u}$. This could arise, e.g., due to efficient pitch-angle scattering dominating the system’s dynamics, and implies the neglect of heat fluxes. Introducing spherical polar coordinates $(\tilde{p}, \theta, \phi)$, with the polar axis orientated in an arbitrary direction, we have
\begin{equation}
    \tilde{\mathbf{p}} = \tilde{p} \left( \cos\phi \sin\theta \hat{\mathbf{x}} + \sin\phi \sin\theta \hat{\mathbf{y}} + \cos\theta \hat{\mathbf{z}} \right),
\end{equation}
and the corresponding Jacobian is
\begin{equation}
\label{eq: J jacobian}
    \left|\frac{\partial\tilde{\mathbf{p}}}{\partial(\tilde{p}, \theta, \phi)}\right| = \tilde{p}^2 \sin \theta.
\end{equation}
The isotropy assumption is then the statement that
\begin{equation}
    f = f(t, \mathbf{r}, \tilde{p}).
\end{equation}
We assume that in this `reduced phase space', phase-space volume is conserved, i.e 
\begin{align}
   \left| \frac{\partial (\mathbf{r}, \tilde{p})}{\partial (\mathbf{r}_0, \tilde{p}_0)}\right| &= \frac{\tilde{p}_0^2}{\tilde{p}^2},
\label{eq: iso J}
\end{align}
where we have made use of \eqref{eq: J jacobian}. Combining \eqref{eq: iso J} with Liouville's theorem~\eqref{eq: Liouville} then implies the conservation of probability 
\begin{align}
\label{eq: iso cont}
    d\mathbf{r} d\tilde{p} \ \tilde{p}^2 f(t, \mathbf{r}, \tilde{p}) &=
    d\mathbf{r}_0 d\tilde{p}_0 \ \tilde{p}_0^2 f_0(\mathbf{r}_0, \tilde{p}_0).
\end{align}

Considering a rapidly isotropised distribution function is equivalent to assuming that the phase space can be described by its average over some fast timescale $t_\text{fast}$ associated with the isotropisation process. In the laboratory frame, such an average is defined by
\begin{equation}
    \langle\cdot\rangle_{t_{\rm fast}}
    \equiv
    \frac{\int_t^{t+t_{\rm fast}}(\cdot)\,dt}
         {\int_t^{t+t_{\rm fast}}dt}.
\end{equation}
However, isotropisation is a microscopic process that occurs in the particle's proper time~$\tau$, which is related to the laboratory time via $dt=\gamma\,d\tau$. We assume that during this process, the particle's motion becomes uniformly randomised as far as its direction is concerned, allowing the proper-time average to be replaced by an average over the solid angle of the peculiar momentum. Therefore,
\begin{equation}
    \label{eq: average}
    \langle\cdot\rangle_{t_{\rm fast}}
    =
    \frac{\int_\tau^{\tau+\tau_{\rm fast}}(\cdot)\gamma\,d\tau}
         {\int_\tau^{\tau+\tau_{\rm fast}}\gamma\,d\tau}
    =
    \frac{\iint(\cdot)\gamma\,\sin \theta \, d \theta d\phi}
         {\iint\gamma\,\sin \theta \, d \theta d\phi}
    =
    \frac{\langle\gamma(\cdot)\rangle_{\theta,\phi}}
         {\langle\gamma\rangle_{\theta,\phi}}.
\end{equation}
Thus, in the relativistic limit, replacing the averaging over fast timescales with angle averaging leads to the appearance of a Lorentz factor inside the angle integrals. As expected, in the non-relativistic limit, where $\gamma \approx 1$, these averages are the same.

Averaging the phase-space flow \eqref{eq: phase-space velocity} via \eqref{eq: average} gives $\langle \dot{\mathbf{r}}\rangle_{t_\text{fast}} = \mathbf{u}$, where we have used \eqref{eq: Isotropic momentum split} and \eqref{eq: Isotropic gamma split}. Consequently, in the reduced phase space, the mapping between $\mathbf{r}$ and $\mathbf{r}_0$ is independent of the peculiar coordinates. The phase-space path $\mathbf{r} = \mathbf{r}(t, \mathbf{r}_0)$ then corresponds to that of a fluid element advected with the bulk velocity $\mathbf{u}$ in three-dimensional position space. The Jacobian \eqref{eq: iso J} therefore factorises into spatial and momentum parts:
\begin{equation}
\label{eq: isotropic J eq almost}
    \left| \frac{\partial (\mathbf{r}, \tilde{p})}{\partial (\mathbf{r}_0, \tilde{p}_0)}\right| = \left|\frac{\partial \mathbf{r}}{\partial \mathbf{r}_0}\right| \frac{\partial \tilde{p}}{\partial \tilde{p}_0}.
\end{equation}
We can calculate the first factor in \eqref{eq: isotropic J eq almost} by returning to \eqref{eq: iso cont}, integrating over the momenta, and recalling the definition of density \eqref{eq: bulk def} to find
\begin{equation}
    d \mathbf{r} \ n(t, \mathbf{r}) = d \mathbf{r}_0 \ n_0(\mathbf{r}_0),
\end{equation}
which implies
\begin{equation}
    \label{eq: cont n}
    \left|\frac{\partial \mathbf{r}_0}{\partial \mathbf{r}}\right| = \frac{n(t, \mathbf{r})}{n_0(\mathbf{r}_0)} \equiv n'(t, \mathbf{r}_0).
\end{equation}
Combining the above expression with \eqref{eq: isotropic J eq almost} and inserting the result into \eqref{eq: iso J} yields a differential equation whose solution is
\begin{equation}
    \label{eq: isotropic p evolution}
    \tilde{p}_0 = \frac{\tilde{p}}{n'^{1/3}}.
\end{equation}

We can now substitute \eqref{eq: isotropic p evolution} into Liouville's theorem \eqref{eq: Liouville} and find the evolution of the distribution function:
\begin{equation}
    \label{eq: isotropic relativistic law}
    f(t, \mathbf{r}, \tilde{p}) = f_0\left( \mathbf{r}_0, \frac{\tilde{p}}{n'^{1/3}} \right).
\end{equation}
Note that the explicit form of $\mathbf{r}(t, \mathbf{r}_0)$ is not required, as the trajectory corresponds to that of a fluid element advected with the bulk velocity $\mathbf{u}$. Hence, any evolution equations obtained by taking moments of \eqref{eq: isotropic relativistic law} are Lagrangian relations with respect~$\mathbf{u}$.

\subsubsection{Equation of State for Isotropic Plasma}
\label{sec: Equation of State for an Isotropic Plasma}
Now that we have found how the distribution function evolves, we can work out the corresponding EoS. In the relativistic regime, pressure is given by
\begin{equation}
    \label{eq: pressure isotropic definition}
    P = \frac{1}{3m} \int d {\tilde{\mathbf{p}}} \, \frac{\tilde{p}^2}{\tilde{\gamma}} \, f.
\end{equation}
Inserting \eqref{eq: isotropic relativistic law} into \eqref{eq: pressure isotropic definition}, integrating over $\theta$ and $\phi$, and changing the momentum integration variable to $\tilde{p}_0$ via \eqref{eq: isotropic p evolution} gives the adiabatic EoS for an isotropic plasma:
\begin{equation}
    \label{eq: Isotropic pressure evolution}
    P = \frac{4 \pi n'^{5/3}}{3m} \int d\tilde{p}_0 \,
    \frac{\tilde{p}^4_0 f_0(\tilde{p}_0)}
    {\sqrt{1 + n'^{2/3} \left( {\tilde{p}_0}/{mc} \right)^2 }}.
\end{equation}
In the non-relativistic regime $\tilde{p} \ll mc$, \eqref{eq: Isotropic pressure evolution} reduces to the familiar scaling $P \propto n^{5/3}$ and in the ultra-relativistic limit $\tilde{p} \gg mc$, we instead obtain $P \propto n^{4/3}$, as expected. In general, however, $P$ depends not solely on density but also on the form of the initial distribution function $f_0(\tilde{p})$. Substituting the Maxwell--Jüttner distribution into \eqref{eq: Isotropic pressure evolution} recovers the Synge gas EoS~\citep{Synge-1957}.

\subsection{Gyrotropy and Parity}
\label{sec: Gyrotropy and Parity}
We now consider a more general case, where the isotropy assumption is relaxed but the distribution function still satisfies two symmetries in the peculiar-momentum coordinate. First, it is gyrotropic, i.e., it has cylindrical symmetry with respect to some vector field, which we call~$\mathbf{B}$. Second, the distribution function has parity symmetry along this vector. We assume that the vector $\mathbf{B}$ is divergence free and frozen into the bulk flow; it therefore satisfies the induction equation
\begin{equation}
\label{eq: induction}
    \frac{\partial \mathbf{B}}{\partial t} = \nabla \times (\mathbf{u} \times \mathbf{B}).
\end{equation}
For example, in a drift-kinetic plasma, $\mathbf{B}$ corresponds to the magnetic field \citep{Kulsrud-1964}. A distribution function with such symmetries can arise in systems where fast processes, such as gyromotion or the bouncing of trapped particles in magnetic mirrors of collisionless plasmas, dominate the dynamics, and implies the neglect of heat fluxes\footnote{Double-adiabatic EoSs hold if certain combinations of parallel-heat-flux terms
vanish at all times. The parity condition is a sufficient condition for this to happen; it is not currently known whether it is also necessary.}.

We adopt a cylindrical coordinate system in the peculiar-momentum space $(\tilde{p}_\perp, \phi, \tilde{p}_\parallel,  \sigma)$, with the polar axis parallel to~$\mathbf{b} \equiv \mathbf{B}/B$:
\begin{equation}
    \tilde{\mathbf{p}} = \tilde{p}_\perp( \cos \phi \hat{\mathbf{x}} + \sin\phi\hat{\mathbf{y}}) +  \sigma \tilde{p}_\parallel \mathbf{b},
\end{equation}
where $\tilde{p}_\parallel \geq 0$ and $\sigma$ takes the values $+1$ or $-1$, labelling the direction of the parallel peculiar momentum. The Jacobian of this change of variables is
\begin{equation}
    \label{eq: gyro J}
    \left|\frac{\partial\tilde{\mathbf{p}}}{\partial( \tilde{p}_\perp, \phi, \tilde{p}_\parallel)}\right| =  \tilde{p}_\perp.
\end{equation}
The gyrotropy and parity assumptions are then equivalent to imposing that $f$ is independent of $\phi$ and~$\sigma$:
\begin{equation}
    \label{eq: sym gyro parity}
    f = f(t, \mathbf{r}, \tilde{p}_\perp, \tilde{p}_\parallel).
\end{equation}
We again assume that phase-space volume in the reduced phase-space is conserved, viz.
\begin{align}
    \left| \frac{\partial (\mathbf{r}, \tilde{p}_\perp, \tilde{p}_\parallel)}{\partial (\mathbf{r}_0, \tilde{p}_{\perp0}, \tilde{p}_{\parallel0})} \right|&= \frac{\tilde{p}_{\perp0}}{\tilde{p}_\perp}
     \label{eq: gyro J 2},
\end{align}
where we have made use of \eqref{eq: gyro J}. Combining Liouville's theorem~\eqref{eq: Liouville} with \eqref{eq: gyro J 2} yields the conservation of probability \eqref{eq: phase-space cont} in our reduced phase space
\begin{align}
    \label{eq: gyro cont}
    d\mathbf{r}d\tilde{p}_\perp d\tilde{p}_\parallel  \ \tilde{p}_\perp  f(t, \mathbf{r}, \tilde{p}_\perp, \tilde{p}_\parallel) &= \nonumber \\
    d\mathbf{r}_0 d\tilde{p}_{\perp0}& d\tilde{p}_{\parallel0}  \ \tilde{p}_{\perp0} f_0(\mathbf{r}_0, \tilde{p}_{\perp0}, \tilde{p}_{\parallel0}).
\end{align}

In analogy to Section \ref{sec: isotropy}, imposing \eqref{eq: sym gyro parity} is equivalent to assuming that the phase space is well described by its average over some fast timescales associated with the two symmetries (e.g., Larmor gyration and bouncing inside magnetic mirrors). In the non-relativistic limit, averaging over these timescales in the laboratory frame is equivalent to an average over $\sigma$ and~$\phi$. In the relativistic case, as in \eqref{eq: average}, the average acquires a weight of~$\gamma$:
\begin{equation}
    \label{eq: average gyro}
    \langle\,\cdot\,\rangle_{t_\text{fast}}
    \equiv \frac{\int_{t}^{t + t_\text{fast}} (\,\cdot\,) dt}{\int_{t}^{t + t_\text{fast}}dt} = 
    \frac{\sum_\sigma\int(\cdot)\gamma\,d\phi}
    {\sum_\sigma\int\gamma\,d\phi}
    = 
    \frac{\langle \gamma(\,\cdot\,)\rangle_{\phi, \sigma}}
    {\langle\gamma\rangle_{\phi,\sigma}}.
\end{equation}

Averaging the phase-space flow \eqref{eq: phase-space velocity} over the fast timescales using \eqref{eq: Isotropic momentum split},  \eqref{eq: Isotropic gamma split}, and \eqref{eq: average gyro} gives $\langle \dot{\mathbf{r}}\rangle_{t_\text{fast}} = \mathbf{u}$, as in Section \ref{sec: isotropy}, implying that in the reduced phase space, the trajectory $\mathbf{r} = \mathbf{r}(t, \mathbf{r}_0)$ again corresponds to that of a fluid element advected with the bulk flow~$\mathbf{u}$. Hence, the Jacobian in \eqref{eq: gyro J 2} factorises into spatial and momentum contributions:
\begin{align}
    \label{eq: gyro J eq almost}
    \left|\frac{\partial (\mathbf{r}, \tilde{p}_\perp, \tilde{p}_\parallel)}{\partial (\mathbf{r}_0, \tilde{p}_{\perp0}, \tilde{p}_{\parallel0})}\right| &= \left|\frac{\partial \mathbf{r}}{\partial \mathbf{r}_0}\right| \left|\frac{\partial( \tilde{p}_\perp, \tilde{p}_\parallel)}{\partial ( \tilde{p}_{\perp0}, \tilde{p}_{\parallel0})}\right|,
\end{align}
where the former can be evaluated using \eqref{eq: cont n}. Then, from \eqref{eq: gyro J eq almost} and \eqref{eq: gyro J 2}, we have
\begin{align}
    \left|\frac{\partial( \tilde{p}_\perp, \tilde{p}_\parallel)}{\partial ( \tilde{p}_{\perp0}, \tilde{p}_{\parallel0})} \right| &= \frac{\tilde{p}_{\perp0}}{\tilde{p}_\perp} n'(t, \mathbf{r}_0).
    \label{eq: p_perp p_parallel}
\end{align}

To determine the evolution of $f$, we require expressions for $\tilde{p}_\perp$ and $\tilde{p}_\parallel$ in terms of their initial values. However, we have obtained only a single constraint~\eqref{eq: p_perp p_parallel}. We therefore assume that, in addition to phase-space volume, the phase-space area perpendicular to $\mathbf{B}$ is conserved. A natural coordinate system for resolving directions parallel and perpendicular to $\mathbf{B}$ is provided by the Clebsch variables $(\alpha, \beta, \ell)$, defined by
\begin{equation}
    \mathbf{B} = \nabla \alpha \times \nabla \beta, \quad \nabla \ell = \mathbf{b}.
\end{equation}
Here, $\alpha$ and $\beta$ label a magnetic-field line, while $\ell$ measures distance along it. The Jacobian of this transformation is
\begin{equation}
    \label{eq: Clebsch J}
    \left|\frac{\partial\mathbf{r}}{\partial(\alpha, \beta, \ell)}\right| = \frac{1}{B(t, \mathbf{r})},
\end{equation}
and the conservation of perpendicular phase-space area can be written as
\begin{align}
    \label{eq: phase-space area cons}
    d\alpha d\beta d \tilde{p}_\perp \ B^{-1}(t, \mathbf{r}) \tilde{p}_\perp  &= d\alpha_0 d\beta_0 d \tilde{p}_{\perp0} \ B_0^{-1}(\mathbf{r}_0) \tilde{p}_{\perp0}
\end{align}
or, equivalently, 
\begin{equation}
     \label{eq: phase-space area cons J}
     \left|\frac{\partial (\alpha, \beta, \tilde{p}_\perp)}{\partial (\alpha_0, \beta_0, \tilde{p}_{\perp0})}\right| = \frac{\tilde{p}_{\perp0}}{\tilde{p}_\perp}\frac{B(t, \mathbf{r})}{B_0(\mathbf{r}_0)} \equiv \frac{\tilde{p}_{\perp0}}{\tilde{p}_\perp} B'(t, \mathbf{r}_0).
\end{equation}
As we discuss in Section \ref{sec: Application to Collisionless, Magnetised Plasmas}, this is just the conservation of $\mu$ in collisionless, magnetised plasmas, or rather the conservation of $\mu$ is how the conservation of perpendicular phase-space area is physically achieved by a plasma. 

From the induction equation \eqref{eq: induction}, we have $\dot{\alpha} = \dot{\beta} = 0$, so $\alpha = \alpha_0$ and $\beta = \beta_0$, i.e., $\alpha$ and $\beta$ remain constant as the field is advected by the flow~$\mathbf{u}$. Hence, \eqref{eq: phase-space area cons J} yields
\begin{equation}
     \label{eq: p_perp int}
     \frac{\partial \tilde{p}_\perp}{\partial \tilde{p}_{\perp 0}} = \frac{\tilde{p}_{\perp0}}{\tilde{p}_\perp} B'.
\end{equation}
Integration of \eqref{eq: p_perp int} yields
\begin{equation}
    \label{eq: p_perp evolve}
    \tilde{p}_{\perp 0} = \frac{\tilde{p}_{\perp}}{\sqrt{B'}},
\end{equation}
and substituting this into \eqref{eq: p_perp p_parallel} gives
\begin{equation}
    \label{eq: p_parallel evolve}
    \tilde{p}_{\parallel0} = \frac{\tilde{p}_\parallel B'}{n'}.
\end{equation}

Finally, we can combine \eqref{eq: p_perp evolve} and \eqref{eq: p_parallel evolve} with Liouville's theorem \eqref{eq: Liouville} to find the evolved distribution function:
\begin{align}
    \label{eq: gyro f evolution}
    f(t, \mathbf{r} , \tilde{p}_\perp, \tilde{p}_\parallel) &= f_0\left( \mathbf{r}_0, \frac{\tilde{p}_\perp}{\sqrt{B'}}, \frac{\tilde{p}_\parallel B'}{n'}\right). 
\end{align}
Again, the time dependence is encoded via $n'(t, \mathbf{r}_0)$,  $B'(t, \mathbf{r}_0)$ and $\mathbf{r}(t, \mathbf{r}_0)$.

\subsubsection{Equations of State for Gyrotropic Plasma}
Similarly to Section \ref{sec: Equation of State for an Isotropic Plasma}, we can now use the evolved distribution \eqref{eq: gyro f evolution} to calculate the corresponding EoSs. For a relativistic plasma, the parallel and perpendicular pressures are defined by
\begin{equation}
    \label{eq: p pressure definitions}
    P_\perp = \frac{1}{2m}\int d {\tilde{\mathbf{p}}} \ \frac{\tilde{p}^2_\perp} {\tilde{\gamma}} f, \quad P_\parallel = \frac{1}{m}\int d{\tilde{\mathbf{p}}} \ \frac{\tilde{p}^2_\parallel} {\tilde{\gamma}} f.
\end{equation}
Inserting \eqref{eq: gyro f evolution} into these equations, integrating over $\sigma$ and $\phi$, and finally changing the integration variables to $\tilde{p}_{\perp0}$ and $\tilde{p}_{\parallel0}$ via \eqref{eq: p_perp evolve} and \eqref{eq: p_parallel evolve} yields the double-adiabatic pressure-evolution equations:
\begin{align}
    P_\perp &=  \frac{2\pi n' B'}{m}\int  d\tilde{p}_{\perp0} \ d\tilde{p}_{\parallel0} \ \nonumber \\
    & \frac{\tilde{p}_{\perp0}^3 f_0(\tilde{p}_{\perp0}, \tilde{p}_{\parallel0}) \ } {\sqrt{1 + B' \left({\tilde{p}_{\perp0}}/{mc} \right)^2 + \left( {n'}/{B'}\right)^2\left({\tilde{p}_{\parallel0}}/{mc} \right)^2}}, 
    \label{eq: full P_perp evolution}\\
    P_\parallel &= \frac{4\pi n'^3}{m B'^2}\int  d\tilde{p}_{\perp0} \ d\tilde{p}_{\parallel0} \ \nonumber 
    \\ &  \frac{\tilde{p}_{\perp0} \tilde{p}_{\parallel0}^2 f_0(\tilde{p}_{\perp0}, \tilde{p}_{\parallel0}) \ } {\sqrt{1 + B' \left({\tilde{p}_{\perp0}}/{mc} \right)^2 + \left( {n'}/{B'}\right)^2\left({\tilde{p}_{\parallel0}}/{mc} \right)^2}}.
    \label{eq: full P_parallel evolution}
\end{align}
These integrals do not, in general, exhibit a single set of simple scalings with $n'$ and $B'$ for arbitrary~$f_0$. However, in the non-relativistic limit, $\tilde{p}_\perp, \tilde{p}_\parallel \ll mc$, \eqref{eq: full P_perp evolution} and \eqref{eq: full P_parallel evolution} reduce to the CGL equations, i.e., $P_\perp \propto n B$ and $P_\parallel \propto n^3/B^2$ \citep{Chew_etal-1997}, for all~$f_0$.

\subsubsection{Modified Pressures}
The complicated dependence of \eqref{eq: full P_perp evolution} and \eqref{eq: full P_parallel evolution} on $n'$ and $B'$ arises due to the presence of the Lorentz factor in the integrand of \eqref{eq: p pressure definitions}. To avoid this issue, one can define the modified pressures
\begin{equation}
    \label{eq: p hat pressure definitions}
    \hat{P}_\perp \equiv \frac{1}{2m}\int d {\tilde{\mathbf{p}}} \ \tilde{p}^2_\perp f, \quad \hat{P}_\parallel \equiv \frac{1}{m}\int d {\tilde{\mathbf{p}}}\ \tilde{p}^2_\parallel f.
\end{equation}
Proceeding as before and inserting \eqref{eq: gyro f evolution} into the definitions above, we find
\begin{align}
    \label{eq: full hat_P_perp evolution}
    \hat{P}_\perp &=  \frac{2\pi n' B'}{m}\int  d\tilde{p}_{\perp0} \ d\tilde{p}_{\parallel0} \  {\tilde{p}_{\perp0}^3 f_0(\tilde{p}_{\perp 0}, \tilde{p}_{\parallel 0})}, \\
    \label{eq: hat_full P_parallel evolution}
    \hat{P}_\parallel &= \frac{4\pi n'^3}{m B'^2}\int  d\tilde{p}_{\perp0} \ d\tilde{p}_{\parallel0} \ \tilde{p}_{\perp0}{\tilde{p}^2_{\parallel 0} f_0(\tilde{p}_{\perp 0}, \tilde{p}_{\parallel 0})}.
\end{align}
The integrals in both \eqref{eq: full hat_P_perp evolution} and \eqref{eq: hat_full P_parallel evolution} can be replaced by the initial values of the modified pressures, $\hat{P}_{\perp0}$ and $\hat{P}_{\parallel0}$. We thus obtain the simple relationships
\begin{equation}
    \label{eq: double adiabatic evolution}
    \hat{P}_\perp' = n'B', \quad \hat{P}_\parallel' = \frac{n'^3}{B'^2},
\end{equation}
where $\hat{P}_\perp' \equiv \hat{P}_{\perp}/\hat{P}_{\perp0}$ and $\hat{P}_\parallel' \equiv \hat{P}_{\parallel}/\hat{P}_{\parallel0}$. These expressions are the relativistic generalisation of the CGL equations. They can alternatively be derived by taking moments of the relativistic Vlasov equation and neglecting heat fluxes, as was first done in \citet{Gedalin_Oiberman-1995}. However, in the fluid momentum equation, it is $P_\perp$ and $P_\parallel$ that appear, not $\hat{P}_\perp$ and $\hat{P}_\parallel$. Hence, in general, \eqref{eq: double adiabatic evolution} does not provide a suitable fluid closure for collisionless relativistic plasmas.

\subsubsection{Application to Collisionless, Magnetised Plasmas}
\label{sec: Application to Collisionless, Magnetised Plasmas}
The theory presented above derives double-adiabatic evolution equations for a fluid system with a divergence-free vector field $\mathbf{B}$ frozen into the bulk flow. The derivation relies on a set of assumptions regarding the symmetries of the distribution function in peculiar-momentum space, as well as the conservation properties of the evolving phase-space density. First, we considered a gyrotropic distribution function and assumed conservation of phase-space area perpendicular to $\mathbf{B}$. Second, we imposed parity symmetry along the direction of $\mathbf{B}$ in peculiar-momentum space, together with conservation of the full phase-space volume.

An example of a system in which such symmetries arise as a result of fast processes is a collisionless magnetised plasma. The fast particle gyration provides the largest frequency in such a system, implying the gyrotropy of the distribution function and the conservation of the first adiabatic invariant \citep{Gardner-1959}
\begin{equation}
    \label{eq: mu}
    \mu = \frac{\tilde{p}_\perp^2}{2B}
\end{equation}
along gyro-averaged particle trajectories. This conservation implies \eqref{eq: p_perp evolve} and provides the physical mechanism whereby the conservation of perpendicular phase-space area \eqref{eq: phase-space area cons} holds. We are currently not aware of a system where such conservation is achieved by different means.

The conservation of $\mu$ can cause particles with sufficiently large pitch angles to become trapped in local minima of the magnetic-field strength (magnetic mirrors) and undergo bounce motion. If such bouncing occurs on timescales much shorter than those of the fluid motion, a second adiabatic invariant,
\begin{equation}
    \label{eq: J}
    J = \oint \tilde{p}_\parallel \, d \ell,
\end{equation}
is conserved, where the integral is taken between the bounce points \citep{Northrop_Teller-1960}. The rapid bouncing enforces parity symmetry along the magnetic-field direction. A direct result of both $\mu$ and $J$ conservation is then \eqref{eq: p_parallel evolve}. The correspondence can be seen more directly by combining \eqref{eq: gyro J 2}, \eqref{eq: Clebsch J}, \eqref{eq: p_perp evolve}, and \eqref{eq: p_parallel evolve} to obtain
\begin{equation}
    \frac{\tilde{p}_\parallel}{\tilde{p}_{\parallel0}} \frac{\partial \ell}{\partial \ell_0} = 1,
\end{equation}
which is consistent with the conservation of $J$ as defined in~\eqref{eq: J}. 

Thus, the conservation of $\mu$ and $J$ in magnetised, collisionless plasmas provides a concrete example in which the assumptions made in Section \ref{sec: Gyrotropy and Parity} hold, and \eqref{eq: full P_perp evolution} and \eqref{eq: full P_parallel evolution} yield accurate EoSs. However, the general approach we have presented here shows that adiabaticity in the fluid-dynamical context arises due to specific symmetries and phase-space conservation properties of the distribution function. It is best highlighted by our isotropic example in Section \ref{sec: isotropy} that individual particles are not required to possess adiabatic invariants for adiabatic EoSs to hold, although that is the case for rapidly gyrating, trapped particles in magnetised collisionless plasmas. Consequently, the requirement of particle trapping is not necessary for the double-adiabatic equations \eqref{eq: full P_perp evolution} and \eqref{eq: full P_parallel evolution} to be valid. The EoSs would hold as long as our assumptions about the symmetries of $f$ and conservation of phase-space volume and area are true. This is why, for example, the CGL equations are found to hold in homogeneous, magnetised, collisionless plasmas, where trapped particles do not exist.

\section{Double-Adiabatic Equations of State for Ultra-relativistic Plasmas}
To make analytical progress in evaluating \eqref{eq: full P_perp evolution} and \eqref{eq: full P_parallel evolution}, we now consider the ultra-relativistic limit, $\tilde{p}_\perp, \tilde{p}_\parallel \gg mc$. For ease of notation, we drop the tildes above the rest-frame quantities. We employ the coordinate transformation $p_\perp = p \sqrt{1 - \xi^2}$ and $p_\parallel = p \xi$, where $\xi$ is the cosine of the pitch angle with respect to the magnetic field. The pressure-evolution equations \eqref{eq: full P_perp evolution} and \eqref{eq: full P_parallel evolution} then simplify to
\begin{align}
    \label{eq: ultra rel P_perp evolution}
    P_\perp &= 2c{\pi}B'^2 \int_{0}^1 d\xi  \int_0^\infty  dp \ \frac{(1-\xi^2)\ p^3 f_0(p, \xi)}{[\xi^2 + (1-\xi^2)B'^3/n'^2]^{1/2}}, \\
    \label{eq: ultra rel P_parallel evolution}
    P_\parallel &= {4c\pi} \frac{n'^2}{B'}\int_{0}^1 d\xi  \int_0^\infty  dp \ \frac{\xi^2\ p^3 f_0(p, \xi)}{[\xi^2 + (1-\xi^2)B'^3/n'^2]^{1/2}}.
\end{align}
The `total' pressure \eqref{eq: pressure isotropic definition}, here equal to $(2P_\perp + P_\parallel)/3$, is given by
\begin{align}
    \label{eq: P def}
     P = \frac{2c \pi}{3} &\frac{n'^2}{B'}\int_{-1}^1 d\xi  \int_0^\infty  dp \nonumber
     \\ & \times p^3[\xi^2 + (1-\xi^2)B'^3/n'^2]^{1/2} f_0(p, \xi),
\end{align}
and has the value
\begin{equation}
\label{eq: P_0 definition}
    P_0 = \frac{2 \pi c}{3}\int_{-1}^1 d\xi  \int_0^\infty  dp \ p^3 f_0(p, \xi)
\end{equation}
at $n' = B' = 1$.

\begin{figure}
\center
\includegraphics[width=0.7\linewidth]{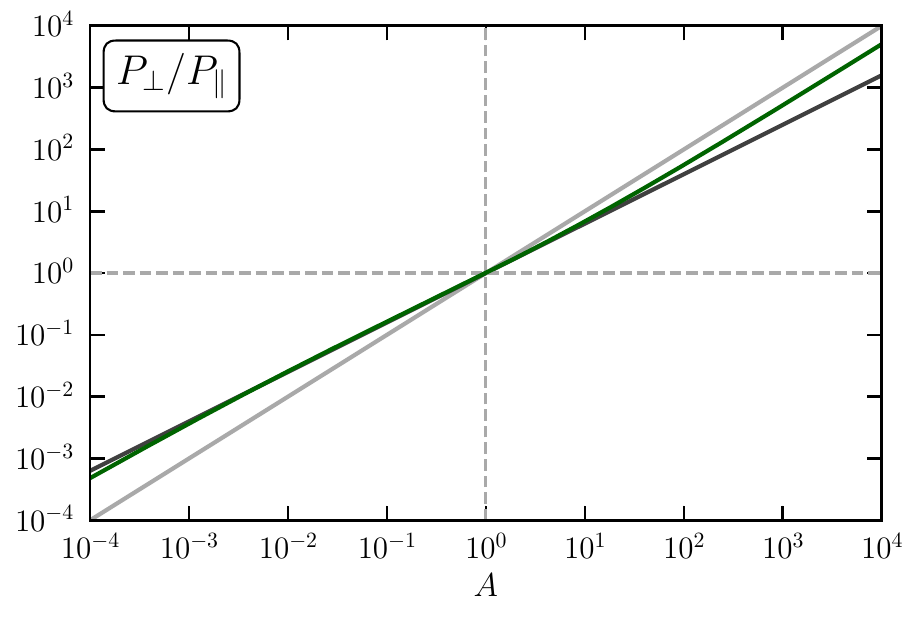}
\caption{Ratio $P_\perp/P_\parallel$ as a function of $A \equiv B'^3/n'^2$ (see \eqref{eq: A}), shown by the green line. The solid light- and dark-grey lines indicate $P_\perp/P_\parallel = A$ and $P_\perp/P_\parallel = A^{4/5}$, respectively. In the limits $P_\perp/P_\parallel \ll 1$ and $P_\perp/P_\parallel \gg 1$, the ratio scales as $P_\perp/P_\parallel \sim A$, whereas close to isotropy it scales as $P_\perp/P_\parallel \sim A^{4/5}$.} 
\label{fig: Anisotropy of I}
\end{figure}

If the initial distribution function is isotropic, i.e., $f_0(p, \xi) = f_0(p)$, the integrals over the momentum magnitude can be replaced by \eqref{eq: P_0 definition}. In this limit, the expressions for $P_\perp$ and $P_\parallel$ are
\begin{align}
\label{eq: pressure perp evolution compressing}
P_\perp' &= B'^2\,
I_\perp\left(\frac{B'^3}{n'^2}\right),
\\
\label{eq: pressure parallel evolution compressing}
P'_\parallel &= \frac{n'^2}{B'}\, I_\parallel\left(\frac{B'^3}{n'^2}\right),
\end{align}
where
\begin{align}
    \label{eq: I_perp full}
    I_\perp(A) &= -\frac{3}{4} \left[\frac{1}{1 - A} +  \frac{(A - 2)}{(1 - A)^{3/2}}\,\operatorname{arsinh}\!\left(\sqrt{\frac{1}{A} - 1}\right)\right], \\
    \label{eq: I_parallel full}
    I_\parallel(A) &= \frac{3}{2} \left[\frac{1}{1 - A} -  \frac{A}{(1 - A)^{3/2}}\,\operatorname{arsinh}\!\left(\sqrt{\frac{1}{A} - 1}\right)\right],
\end{align}
and 
\begin{equation}
    A \equiv \frac{B'^3}{n'^2}.
\end{equation}
Here we have also defined $P_\perp' \equiv P_\perp / P_{0}$ and $P_\parallel' \equiv P_\parallel / P_{0}$. In the case $A > 1$, the identity $\operatorname{arsinh}\xi = -i\arcsin(i\xi)$ can be used to rewrite \eqref{eq: I_perp full} and \eqref{eq: I_parallel full} in an explicitly real form. Key dimensionless physical parameters, viz., $\beta_\perp \equiv 8\pi P_\perp/B^2$, $\beta_\parallel \equiv 8\pi P_\parallel/B^2$, and the pressure anisotropy $P_\perp/P_\parallel$, depend only on the parameter $A$. For example, the latter is
\begin{equation}
    \label{eq: A}
    \frac{P_\perp}{P_\parallel} = A \frac{I_\perp (A)}{I_\parallel(A)},
\end{equation}
which we plot in Figure \ref{fig: Anisotropy of I}.

The relations \eqref{eq: pressure perp evolution compressing} and \eqref{eq: pressure parallel evolution compressing} are the double-adiabatic EoSs for an ultra-relativistic collisionless magnetised plasma with an initially isotropic distribution function. As the point $t=0$ is arbitrary, the above EoS holds as long as $f$ is isotropic at some point in time. An appealing property of these relations is their independence of the specific form of $f_0(p)$, provided that most particles contributing to the pressure tensor are ultra-relativistic. If we impose isotropic evolution, $P_\perp = P_\parallel \equiv P$, then setting \eqref{eq: ultra rel P_perp evolution} and \eqref{eq: ultra rel P_parallel evolution} equal to each other yields $B'^3 = n'^2$. Inserting this back into either of the pressure equations gives the expected result for an isotropic ultra-relativistic plasma:
\begin{equation}
    P'_\perp = P'_\parallel =P' = n'^{4/3}.
\end{equation}

It is useful to consider the EoSs \eqref{eq: pressure perp evolution compressing} and \eqref{eq: pressure parallel evolution compressing} in three key asymptotic limits: 
\begin{enumerate}
  \item large perpendicular pressure, $P_\perp \gg P_\parallel \implies A\gg 1$:
    \begin{equation}
        \label{eq: P perp big}
      P_\perp' =  \frac{3\pi}{8}n' B'^{1/2}, \quad P_\parallel' =\frac{3 \pi }{4} \frac{n'^3}{B'^{5/2}};
    \end{equation}
  \item almost isotropic plasma, $P_\perp \approx P_\parallel \implies |A - 1| \ll 1$:
    \begin{equation}
      \label{eq: P isotropic limit}
      P_\perp' = (n'B')^{4/5}, \quad P_\parallel' = \left(\frac{n'^{3}}{B'^{2}}\right)^{4/5};
    \end{equation}
  \item large parallel pressure,  $P_\perp \ll P_\parallel \implies A\ll 1$:   
  \begin{equation}
  \label{eq: P perp small}
      P_\perp' = \frac{3}{4} B'^2\ln \left( \frac{n'^2}{B'^3}\right), \quad P_\parallel' = \frac{3}{2}\frac{n'^2}{B'}.
    \end{equation}
\end{enumerate}
The asymptotic limits of significant pressure anisotropy agree with the preliminary, heuristic discussion in \cite{Gedalin_Oiberman-1995}, except for the logarithmic correction in \eqref{eq: P perp small}, which arises from a careful treatment of the limit $P_\perp \ll P_\parallel$. In such highly anisotropic regimes, the EoSs can be derived straightforwardly from scaling arguments. For example, when $P_\perp \gg P_\parallel$, we have $\gamma \propto p_\perp$. From the conservation of $\mu$ \eqref{eq: mu}, the perpendicular momentum scales as $p_\perp \propto B^{1/2}$. Conservation of $J$ \eqref{eq: J} implies $p_\parallel \propto n/B$. Substituting these scalings into \eqref{eq: p pressure definitions}, we find $P_\perp \propto n B^{1/2}$ and $P_\parallel \propto n^{3}/B^{5/2}$. A similar argument applies in the limit $P_\parallel \gg P_\perp$, where $\gamma \propto p_\parallel$, yielding $P_\perp \propto B^2$ and $P_\parallel \propto n^2/B$. This discussion also implies that the EoSs obtained in the asymptotic limits $P_\perp \gg P_\parallel$ and $P_\parallel \gg P_\perp$ hold even if $f_0$ is not isotropic, as found in \cite{Gedalin_Oiberman-1995} and as can be shown directly from \eqref{eq: ultra rel P_perp evolution} and \eqref{eq: ultra rel P_parallel evolution}.

\subsection{Generalised Adiabatic Indices}
\begin{figure}
\center
\includegraphics[width=\linewidth]{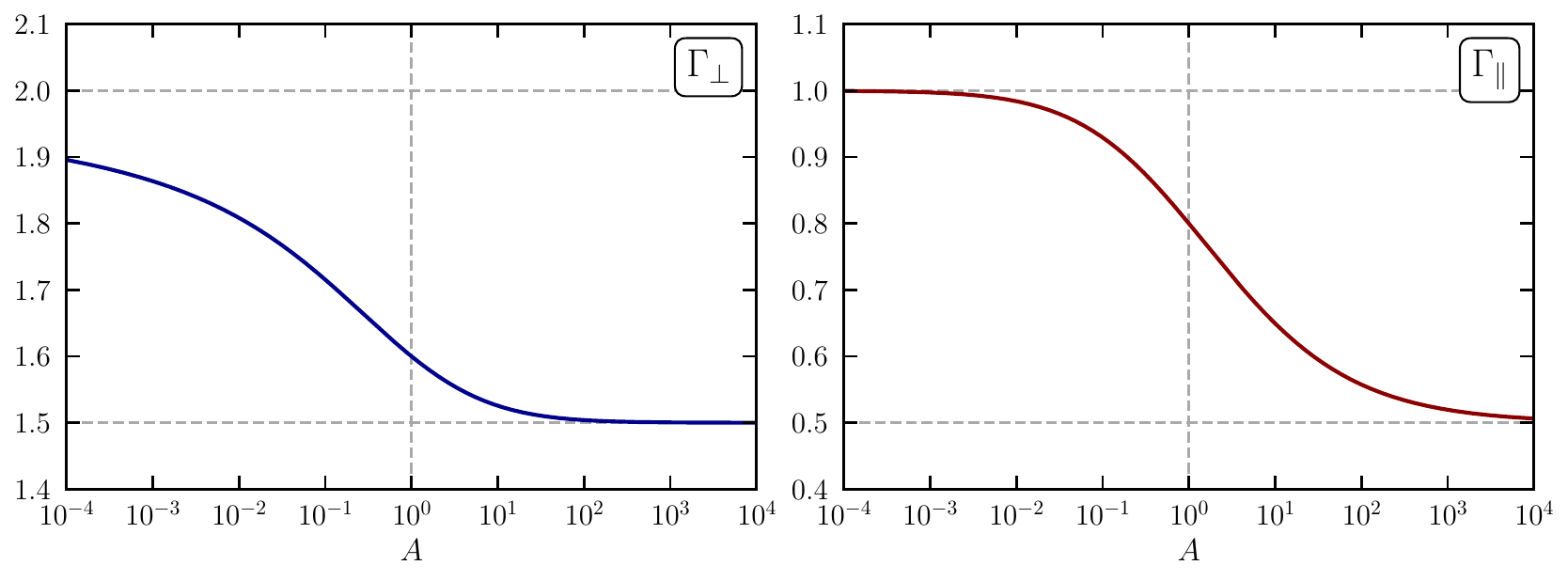}
\caption{Generalised adiabatic indices $\Gamma_\perp$ (left) and $\Gamma_\parallel$ (right), defined in Section \ref{sec: Adiabatic Indices}, as functions of the pressure anisotropy parameter $A = B'^3/n'^2$. The pressure-evolution equations written in terms of these generalised adiabatic indices are \eqref{eq: adiabatic indices}. The dashed vertical lines correspond to the point $A= 1$, i.e., isotropic plasma.} 
\label{fig: Gamma plots}
\end{figure}

\label{sec: Adiabatic Indices}
For ease of use in plasma fluid models, we express $I_\perp$ and $I_\parallel$ in the form of power laws, $I_\perp(A) = A^{\Gamma_\perp -2}$ and $I_\parallel(A) = A^{\Gamma_\parallel -1}$, where $\Gamma_\perp$ and $\Gamma_\parallel$ are, in general, functions of $A$. The pressure-evolution equations \eqref{eq: pressure perp evolution compressing} and \eqref{eq: pressure parallel evolution compressing} are then
\begin{align}
    \label{eq: adiabatic indices}
P_\perp' = B'^2\,
\left(\frac{n'^2}{B'^3}\right)^{2-\Gamma_\perp}, \quad 
P'_\parallel = \frac{n'^2}{B'}\left(\frac{n'^2}{B'^3}\right)^{1-\Gamma_\parallel}.
\end{align}
We refer to $\Gamma_\perp$ and $\Gamma_\parallel$ as the perpendicular and parallel generalised adiabatic indices, respectively, and plot them as a function of $A$ in Figure~\ref{fig: Gamma plots}. The values of  $\Gamma_\perp$ and $\Gamma_\parallel$ in the asymptotic limits of $A$ can be found by comparing \eqref{eq: adiabatic indices} with \eqref{eq: P perp big}, \eqref{eq: P isotropic limit}, and \eqref{eq: P perp small}. Close to isotropy, $|A-1|\ll 1$, $(\Gamma_\perp, \Gamma_\parallel) = (8/5, 4/5)$ and for $A\gg 1$, $(\Gamma_\perp, \Gamma_\parallel) = (3/2, 1/2)$. When $A\ll 1$, $\Gamma_\parallel = 1$, but $\Gamma_\perp$ diverges due to the logarithmic dependence of $I_\perp(A)$ on $A$ in this limit. 

\subsection{Eulerian Formulation}
We have so far worked in a Lagrangian framework with respect to the bulk velocity $\mathbf{u}$. Numerical implementation of \eqref{eq: ultra rel P_perp evolution} and \eqref{eq: ultra rel P_parallel evolution}, however, is often simpler when the EoSs are written in their Eulerian form:
\begin{equation}
    \label{eq: E P}
    \frac{d}{dt} \left(\frac{P_\perp}{B^2 I_\perp(A)}\right) = 0, \quad  \frac{d}{dt} \left(\frac{P_\parallel B}{n^2 I_\parallel(A)}\right) = 0.
\end{equation}
Here $d/dt \equiv \partial/\partial t + \mathbf{u} \cdot \nabla$ is the convective time derivative, and $A = A(P_\perp/P_\parallel)$ is related implicitly to the pressures by \eqref{eq: A}. The absence of primed quantities makes these equations straightforward to implement in fluid simulations and facilitates the inclusion of additional non-ideal effects, such as cooling. 

\section{PIC Simulations of Ultra-Relativistic Plasma in a Compressing Box}

To test the double-adiabatic EoSs that we have derived, we carry out fully kinetic, two-dimensional PIC simulations using the relativistic code \texttt{OSIRIS} \citep{Fonseca_etal-2002}. We employ a set-up in which an electron--positron plasma is compressed perpendicularly to a straight uniform magnetic field of initial strength $B_0$. As a result, both the density and magnetic field increase while maintaining $B' = n'$, meaning that, in this set-up, $A = n'$. We initialise each species with a Maxwell--Jüttner distribution
\begin{equation}
    \label{eq: MJ}
    f(\mathbf{p}) \propto e^{-\gamma m_e c^2 /T_0},
\end{equation}
with temperature $T_0 = 10 \, m_e c^2$, and density $n_0$ such that $\beta_0 = 8 \pi n_0T_0/B_0^2= 1$ for both species. We choose these parameters in order to postpone the onset of pressure-anisotropy-driven kinetic instabilities \citep{Chandrasekhar_etal-1958, Hasegawa-1969, Southwood_Kivelson-1993, Bott_etal-2024}, which will eventually break the double-adiabatic evolution. The full details of our simulations are discussed in Appendix \ref{Appendix: Details of PIC Simulations}.

\begin{figure}
    \centering
    \begin{minipage}[t]{0.499\linewidth}
        \centering
        \includegraphics[width=\linewidth]{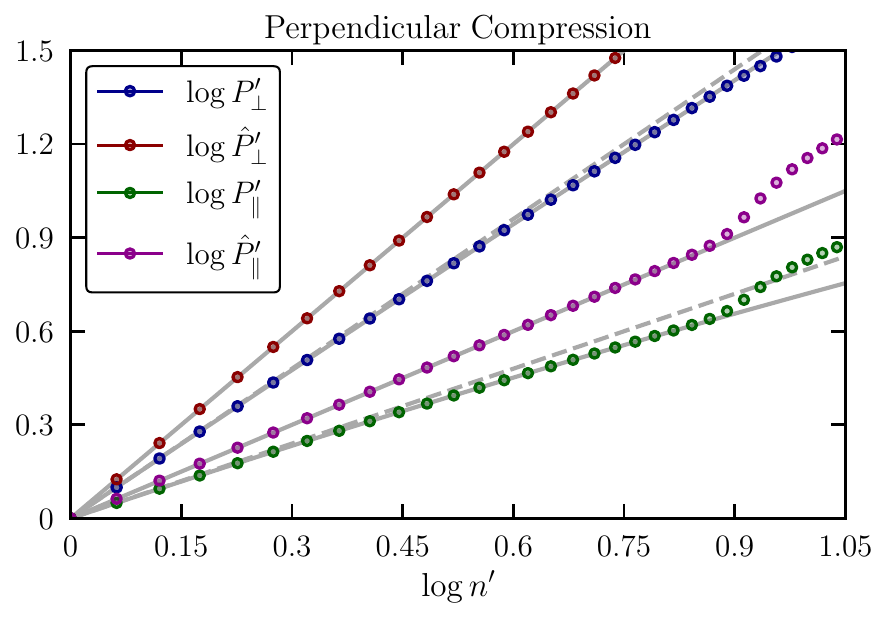}
    \end{minipage}\hfill
    \begin{minipage}[t]{0.499\linewidth}
        \centering
        \includegraphics[width=\linewidth]{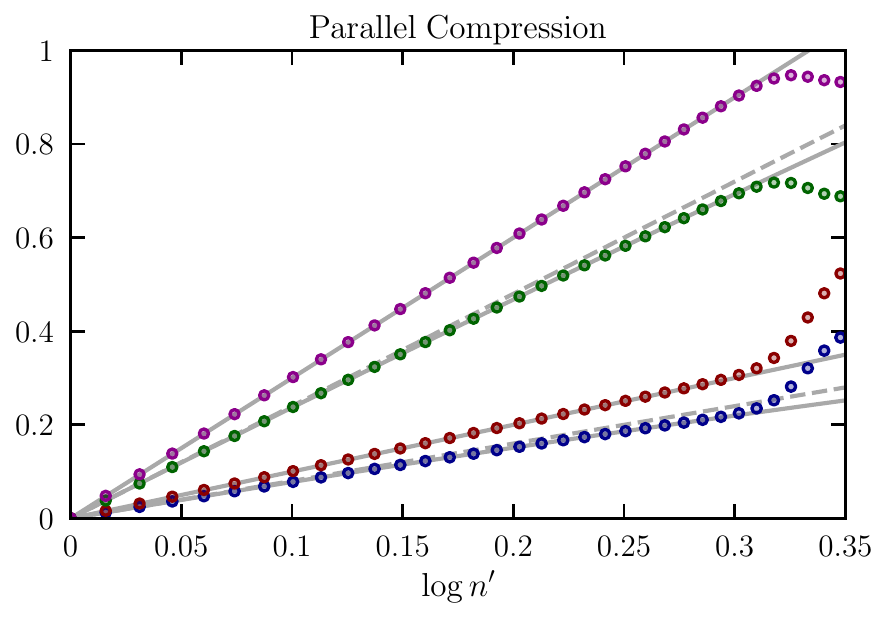}
    \end{minipage}
\caption{The true \eqref{eq: p pressure definitions} and modified \eqref{eq: p hat pressure definitions} pressures, calculated from PIC simulations of plasma undergoing compression perpendicular (left panel) and parallel (right panel) to the magnetic field, plotted as functions of the normalised density $n'$. In the perpendicular compression case, $ B' = n' $, while in the parallel case $ B' = 1 $. The theoretical predictions given by \eqref{eq: pressure perp evolution compressing}, \eqref{eq: pressure parallel evolution compressing}, and \eqref{eq: double adiabatic evolution} are shown as solid grey lines. In the left panel, the grey dashed lines correspond to $ P_\perp' = n'^{8/5} $ and $ P_\parallel' = n'^{4/5} $, while in the right panel to $ P_\perp' = n'^{4/5} $ and $ P_{\parallel}' = n'^{12/5}$. At larger compressions, a clear departure from double-adiabatic theory is observed due to the excitation of the relevant pressure-anisotropy-driven kinetic instabilities.} 
\label{fig: Numerical perp flow Gamma}
\end{figure}

We observe that, as a result of the compression, both the true and modified pressures increase with time as predicted. Plots of these quantities obtained from our simulations, as functions of $n'$, are shown in the left panel of Figure~\ref{fig: Numerical perp flow Gamma}. We find excellent agreement with our theoretical predictions. The evolution of $P_\perp$ and $P_\parallel$ satisfies \eqref{eq: pressure perp evolution compressing} and \eqref{eq: pressure parallel evolution compressing} with $B' = n'$. Figure~\ref{fig: Numerical perp flow Gamma} shows that at values of $n'$ close to unity, i.e., for small anisotropies, $P'_\perp = n'^{8/5}$ and $P'_\parallel = n'^{4/5}$, as expected from \eqref{eq: P isotropic limit}, until $n' \approx 3.2$, when the assumption $P_\perp \sim P_\parallel$ no longer holds and the full pressure equations are required. Our simulations also show that the evolution of $\hat{P}'_\perp$ and $\hat{P}'_\parallel$ is precisely captured by~\eqref{eq: double adiabatic evolution}.

Eventually, when $\Delta \sim 1/\beta_\perp$, around $n' \approx 6.3$, we begin to observe an increase in magnetic energy, indicating the onset of the relativistic mirror instability \citep{Galishnikova_etal-2023}. As a result of the microscale fluctuations in the magnetic field excited by the instability, double-adiabatic evolution breaks down at $n' \approx 7.9$ due to $\mu$ no longer being conserved. Hence, the pressures and modified pressures cease to satisfy \eqref{eq: pressure perp evolution compressing}, \eqref{eq: pressure parallel evolution compressing}, and \eqref{eq: double adiabatic evolution}. In future work, we will further study the late-time evolution of these systems, where the onset of kinetic instabilities becomes important.

To check the general form of the ultra-relativistic pressure-evolution equations \eqref{eq: pressure perp evolution compressing} and \eqref{eq: pressure parallel evolution compressing}, we also ran simulations in which the plasma was compressed parallel to the magnetic field. The simulation results are shown in the right panel of Figure~\ref{fig: Numerical perp flow Gamma}. During such compression, $n'$ still increases with time, but $B'$ remains fixed at unity. We again found good agreement with our double-adiabatic evolution equations, observing initial growth of the pressures given by $P_\perp' = n'^{4/5}$ and $P_\parallel' = n'^{12/5}$, as predicted by \eqref{eq: P isotropic limit}. In this arrangement, $A = 1/n'^2$, and so during the compression a negative pressure anisotropy arises. As a result, we observe the onset of the firehose instability at $\Delta \sim -1/\beta_\parallel$, occurring at $n' \approx 1.9$. Once the firehose fluctuations become sufficiently large, they begin to scatter particles, thereby double-adiabatic evolution no longer holds due to a breaking of $\mu$ conservation. This behaviour is evident in Figure~\ref{fig: Numerical perp flow Gamma} at $n' \approx 2.1$, where the pressure and modified-pressure evolutions depart from \eqref{eq: pressure perp evolution compressing}, \eqref{eq: pressure parallel evolution compressing}, and~\eqref{eq: double adiabatic evolution}.

\section{Discussion}
\label{sec: Discussion}
In this paper, we have shown that phase-space symmetries naturally lead to adiabatic evolution equations for the pressure(s). We did so by treating the distribution function as the density of a six-dimensional, incompressible fluid and demonstrating that imposing such symmetries leads to a self-similar evolution of the distribution function. Taking the appropriate moments of the evolved distribution function then yields adiabatic evolution equations whose form, in general, depends on the initial distribution function.

For a distribution function that is isotropic in the peculiar momentum, such as that of a plasma with efficient pitch-angle scattering, we found an adiabatic EoS for the pressure, which reduces to $P \propto n^{5/3}$ and $P \propto n^{4/3}$ in the non-relativistic and ultra-relativistic limits, respectively. We also considered a distribution whose peculiar-momentum dependence is gyrotropic with respect to a vector field, $\mathbf{B}$ (the magnetic field), frozen into the bulk flow and symmetric under parity along that direction. We found that such a system satisfies double-adiabatic equations. A natural realisation of this scenario is a collisionless, magnetised plasma, in which particles gyrate rapidly around the magnetic field and undergo fast bounce motion in magnetic mirrors. In this case, the double-adiabatic evolution equations govern the parallel and perpendicular pressures, and our symmetry assumptions are consistent with the conservation of the first and second adiabatic invariants, $\mu$ and $J$.

Through developing this symmetry-based formalism, we extended the double-adiabatic evolution equations of a collisionless plasma to the relativistic regime. For an initially isotropic, ultra-relativistically hot plasma, we found $P_\perp \propto B^2 I_\perp$ and $P_\parallel \propto n^2 I_\parallel/B$, where $I_\perp$ and $I_\parallel$ are functions of the compression ratio $B'^3/n'^2$, or, equivalently, of the pressure anisotropy. We evaluated these functions in three asymptotic limits and derived the corresponding pressure-evolution equations. Close to isotropy, $P_\perp \propto (nB)^{4/5}$ and $P_\parallel \propto (n^{3}/B^{2})^{4/5}$. When $P_\perp \gg P_\parallel$, we obtained $P_\perp \propto n B^{1/2}$ and $P_\parallel \propto n^3/B^{5/2}$. In the opposite limit, $P_\parallel \gg P_\perp$, we found $P_\perp \propto B^2 \ln(n'^2/B'^3)$ and $P_\parallel \propto n^2/B$. We also expressed our EoSs in a form ready to be implemented in numerical fluid simulations. As the `initial' time is arbitrary, these EoSs hold if the plasma was, at some point in time, isotropic. We tested these pressure-evolution equations numerically using compressing-box PIC simulations and found excellent agreement with the theoretical predictions. 

A limitation of the double-adiabatic EoSs presented here is that, for collisionless plasmas, the distribution function is only gyrotropic if the coordinate $\mathbf{r}$ refers to the gyrocentre of the particle. In this case, the statement that the averaged particle velocity is equal to the bulk velocity, $\langle \dot{\mathbf{r}}\rangle_{t_\text{fast}} = \mathbf{u}$, neglects magnetic drifts and is therefore valid only in the limit of vanishing gyroradius. Additionally, double adiabaticity requires neglect of heat fluxes and collisions. Incorporating such non-ideal and finite-gyroradius effects would therefore require corrections to the pressure-evolution equations derived here. 

Our derivation was also restricted to plasmas with a non-relativistic bulk flow. An extension to the fully relativistic regime is possible and yields the same EoSs, with the time coordinate replaced by the proper time of the bulk flow \citep{Wierzchucka_etal-2026}. In this regime, $n$ and $B$ denote, respectively, the density and magnetic-field strength in the comoving frame. Constructing a covariant Lagrangian description of phase-space evolution is mathematically involved, owing, e.g., to the need for relativistic Clebsch variables. It turns out to be considerably easier to derive these results from the relativistic drift-kinetic equation \citep{Wierzchucka_etal-2026}.

The physical set-up that we consider here provides a natural starting point for a study of driven kinetic processes, such as the excitation of the firehose and mirror instabilities, in high-energy plasmas. Existing studies suggest that relativistic effects introduce order-unity modifications to the instability thresholds \citep{Galishnikova_etal-2023, Zhdankin_etal-2023}, yet many aspects of their non-linear evolution and saturation in the relativistic regime remain poorly understood, in particular, how driven relativistic mirror and firehose instabilities evolve in the non-linear regime. The relativistic EoSs derived in this work already indicate departures from non-relativistic behaviour. For example, the perpendicular-pressure evolution~\eqref{eq: pressure perp evolution compressing} implies that, unlike in the non-relativistic regime, $\beta_\perp = 8\pi P_\perp / B^2$ decreases under compression perpendicular to the magnetic field. As a result, significantly larger compressions are required to trigger instabilities associated with positive pressure anisotropy, suggesting that relativistic collisionless plasmas could be more stable to anisotropy-driven instabilities than their non-relativistic counterparts.

Our theory of double-adiabatic evolution is relevant to other processes occurring in relativistic plasmas, such as the emission of synchrotron radiation. Recent studies have found that synchrotron radiation generates a negative pressure anisotropy in a cooling plasma, exciting the firehose and maser instabilities \citep{Bilbao_Silvia-2023, Zhdankin_etal-2023, Bilbao_etal-2024}. When the firehose instability interacts with the synchrotron cooling instability \citep{Simon_Axford-1967}, a two-phase medium forms \citep{Wierzchucka_etal-2026b}. One of the phases is high-$\beta$, and its dynamics are dominated by the firehose instability, which keeps the pressure anisotropy pinned to the firehose marginality condition. The second phase is low-$\beta$, and therefore no firehose fluctuations exist within it. The EoSs found in this work have enabled an effective description of such a medium using a fluid model in which the two phases have different effective adiabatic indices \citep{Wierzchucka_etal-2026b}. The determination of effective adiabatic indices through double-adiabatic EoSs is also important for jump conditions of magnetised, collisionless shocks \citep{Sironi_Spitkovsky-2009, Sironi_etal-2013, Bret_Narayan-2018}.

This work may be important for the study of plasmoid dynamics in magnetic reconnection in both astrophysical \citep{Oieroset_etal-2001, Retino_etal-2007} and laboratory \citep{Hare_etal-2017, Totorica_etal-2017} settings. Reversing magnetic-field configurations are unstable and collapse to form current sheets \citep{Furth_etal-1963}. At sufficiently high Lundquist number, these sheets fragment into secondary magnetic islands, or plasmoids \citep{Loureiro_etal-2007, Uzdensky_etal-2010}. In the relativistic regime, magnetic reconnection has been observed to generate extended non-thermal distributions and accelerate particles \citep{Sironi_Spitkovsky-2014, Werner_etal-2015}. Several mechanisms have been proposed to explain this acceleration and the formation of power-law tails, including plasmoid mergers and Fermi-type processes \citep{Drake_etal-2006, Li_etal-2018}. Double-adiabatic models based on the conservation of $\mu$ and $J$ have been used to describe Fermi acceleration, although primarily in the non-relativistic regime \citep{Le_etal-2009, Montag_etal-2017}. PIC simulations of relativistic reconnection have also shown that conservation of the first two adiabatic invariants can energise particles within compressing plasmoids during the `secondary acceleration phase', following initial acceleration in current sheets \citep{Petropoulou_Sironi-2018, Hakobyan_etal-2021}. The double-adiabatic EoSs derived here offer a useful model for describing particle energisation within plasmoids formed during relativistic reconnection.

To conclude, in this paper, we have used a symmetry-based approach to extend the double-adiabatic equations to the relativistic regime, providing a ready-to-implement fluid description for collisionless, magnetised relativistic plasmas. The generality of the formalism suggests that it may also be applicable to other systems satisfying similar assumptions, such as the presence of a frozen-in vector field.

Note: right before the submission of this manuscript, we became aware of similar results that were obtained concurrently and independently by \cite{Ley_etal-2026} and submitted simultaneously with this paper. There are two key differences between our work and that of \cite{Ley_etal-2026}. First, they focus on deriving the adiabatic evolution of $f$ using the (homogeneous) drift-kinetic equation, as opposed to phase-space symmetries. Secondly, while we show that the relativistic double-adiabatic equations \eqref{eq: pressure perp evolution compressing} and \eqref{eq: pressure parallel evolution compressing} hold for any distribution function that was at some point in time isotropic, \cite{Ley_etal-2026} limit themselves to an initial Maxwell--Jüttner equilibrium. Additionally, also shortly before submission, we learned about an otherwise unpublished PhD thesis by P.~Simeon \citep{Simeon-2014} containing a preliminary calculation of adiabatic indices in relativistic, collisionless plasmas. In one chapter, the expressions \eqref{eq: full P_perp evolution} and \eqref{eq: full P_parallel evolution} are derived, and the integrals are evaluated in the limit close to the initial state, i.e., $B' \approx 1$ and $n' \approx 1$, yielding the EoSs~\eqref{eq: P isotropic limit}. 

\section*{Acknowledgments}
We thank A. F. A. Bott, P. G. Ivanov, E. J. Kolmes, M. Lyutikov, and M. L. Nastac for useful discussions. We are especially thankful to F. Bacchini for giving us access to the compressing-box module in the Zeltron code, which inspired us to pursue this project. Simulations were performed at Deucalion (Portugal), funded by FCT MAPs 2: Masers in Astrophysical Plasmas I.P. project 2025.00058.CPCA.A3, and at MareNostrum 5 (Spain), funded by EuroHPC-JU extreme scale access project KEEP: Kinetic Electrodynamics in Extreme Plasmas (EHPC-EXT-2025E02-066). A.W. was supported by a Clarendon Scholarship and the Merton College Tira Wannamethee Scholarship. P.J.B. was supported by a Leverhulme–Peierls Fellowship. A.G.R.T. was supported by US DOE grant DE-SC0024562 and NNSA Center of Excellence DE-NA0003869. The work of A.A.S. and P.J.B. was supported in part by the UK STFC grant ST/W000903/1. A.A.S. and A.W. were also supported in part by the Simons Foundation via the Simons Investigator Award to A.A.S. This research was supported in part by grant NSF PHY-2309135 to the Kavli Institute for Theoretical Physics (KITP).

\section*{Declaration of Interests}
The authors report no conflict of interest.

\appendix

\section{Details of PIC Simulations}
\label{Appendix: Details of PIC Simulations}
\begin{figure}
\center
\includegraphics[width=0.7\linewidth]{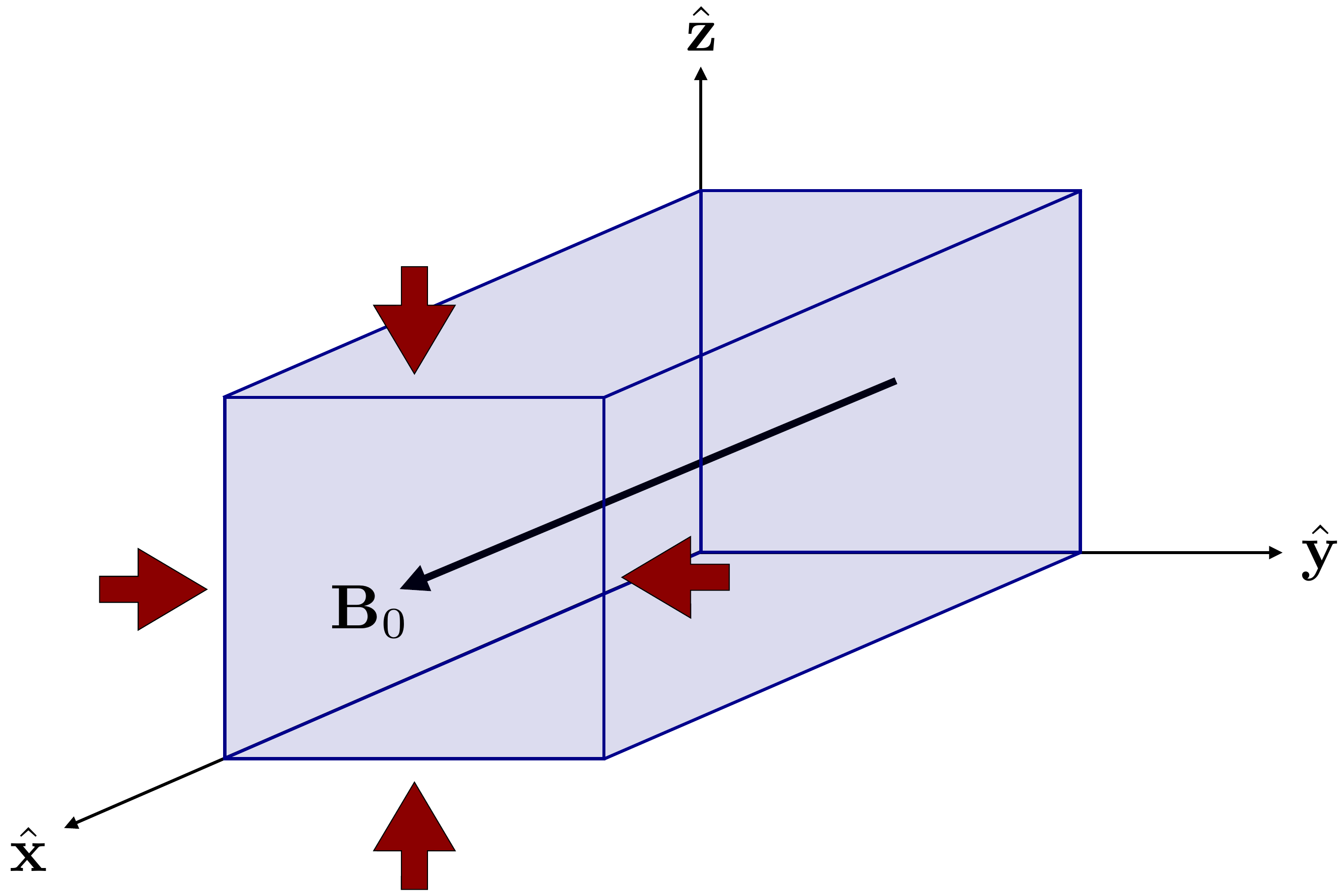}
\caption{Our simulation set-up. The background magnetic field is $\mathbf{B}_0 = B_0 \hat{\mathbf{x}}$, the box is compressed in the $z$ and $y$ directions. Only variations in the $(x,y)$ plane are allowed in the simulation.} 
\label{fig: compressing box diagram}
\end{figure}

To investigate numerically the double-adiabatic EoSs of relativistic plasmas, we use \texttt{OSIRIS} \citep{Fonseca_etal-2002}, a fully relativistic, three-dimensional, highly parallelised PIC code. The PIC numerical method works by solving the equations of motion for charged particles subject to the Lorentz force and by calculating the self-consistent fields on a grid. It is commonly used in plasma physics because of its ability to model particle interactions accurately, thereby allowing the capture of effects beyond the fluid approximation.

We initialised our simulations with a magnetic field $\mathbf{B}_0 = B_0\hat{\mathbf{x}}$ and allowed spatial variations only in the $(x,y)$ plane. This choice reflects the fact that kinetic processes, such as the mirror instability, produce fluctuations with wave vectors both parallel and perpendicular to the magnetic field. Our plasma consisted of electrons and positrons of mass $m_e$. At the start of the simulation, each species had a Maxwell--Jüttner distribution \eqref{eq: MJ} with temperature $T_0 = 10 \, m_e c^2$ and an associated (ultra-relativistic) gyration frequency $\Omega_{e0} = e B_0/ 3 \theta_0 m_e c$, where $\theta_0 \equiv T_0/m_ec^2$. All times and distances were normalised to the initial gyration period $\Omega_{e0}^{-1}$ and the Larmor radius $\rho_{e0} = c/\Omega_{e0}$, respectively. We set the initial densities of the positrons and electrons to $n_0 = \omega_{pe}^2 m_e/ 4\pi e^2$, where $\omega_{pe}$ is the (non-relativistic) plasma frequency. As we are interested in purely collisionless dynamics undisturbed by instabilities, we used $\beta_0 = 8\pi n_0 T_0/B_0^2 = 1$, corresponding to $B_0 = 4.47 \ \omega_{pe} m_e c/e$ and $\omega_{pe}/\Omega_{e0} = 6.7$.

Since the EoSs \eqref{eq: ultra rel P_perp evolution} and \eqref{eq: ultra rel P_parallel evolution} express perpendicular and parallel pressures in terms of the density and magnetic-field strength, we set up a large-scale compressing flow in our simulations to introduce controlled variations in $B'$ and $n'$. The standard PIC treatment of a background, non-relativistic compressing flow, described in \cite{Sironi_Narayan-2015}, is to solve the PIC equations in a frame comoving with the compressing flow. We implemented this method in \texttt{OSIRIS} by modifying the existing shearing-box algorithm \citep{Inchingolo_etal-2018}. For a general compression, the laboratory coordinate $\mathbf{x}$ is related to the comoving coordinate $\mathbf{x'}$ by $\mathbf{x} = \mathbf{L} \mathbf{x'}$, where
\begin{equation}
\label{eq: L def}
    \mathbf{L} = \frac{\partial \mathbf{x}}{\partial \mathbf{x'}} = 
    \begin{pmatrix}
    q_x & 0 & 0 \\
    0 & q_y & 0 \\
    0 & 0 & q_z
    \end{pmatrix},
\end{equation}
and $q_x$, $q_y$, and $q_z$ are time-dependent compression coefficients. To model a compression perpendicular to the magnetic field, we set
\begin{equation}
    q_x = 1, \quad q_y = q_z = (1 + S t)^{-1},
\end{equation}
where $t$ is time and $S$ is the compression rate. In order for the magnetic field to vary slowly compared with the gyration time, we choose $S/\Omega_{e0} = 10^{-2}$. As a result of this compression, the magnetic field and density satisfy $B' = n' = (1 + St)^2$. Our set-up is summarised in Figure~\ref{fig: compressing box diagram}. We also ran simulations in which the plasma was compressed parallel to the guide field, i.e.,
\begin{equation}
    q_x = (1 + S t)^{-1}, \quad q_y = q_z = 1,
\end{equation}
with all other parameters identical to those of the perpendicular-compression simulation. In this case, $n' = 1 + St$, whereas the magnetic field remains constant during the compression.

Our simulations used a spatial resolution $\Delta x = 0.036 \, \rho_{e0}$ in each direction and a time step of $\Delta t = 0.0055 \, \Omega_{e0}^{-1}$. This ensured that the Courant–Friedrichs–Lewy condition in two dimensions, $c \Delta t / \Delta x < 1/\sqrt{2}$, was satisfied throughout the entire duration of the compression and that Larmor-scale motions were properly resolved. The domain size was $L_x \times L_y = 20.1 \rho_{e0} \times 20.1 \rho_{e0}$, with a corresponding number of cells $N_x \times N_y = 552 \times 552$. We imposed periodic boundary conditions in both directions and used $342$ particles per cell with cubic particle shape interpolation.

\bibliographystyle{jpp}
\bibliography{references}

@incollection{Kunz_etal-2022,
	title = {Plasma {physics} of the {intracluster} {medium}},
	isbn = {978-981-16-4544-0},
	url = {https://link.springer.com/rwe/10.1007/978-981-16-4544-0_125-1},
	doi = {10.1007/978-981-16-4544-0_125-1},
	urldate = {2026-03-25},
	booktitle = {Handbook of {X}-ray and {Gamma}-ray {Astrophysics}},
	publisher = {Springer, Singapore},
	author = {Kunz, M. W. and Jones, T. W. and Zhuravleva, I.},
	year = {2022},
	doi = {10.1007/978-981-16-4544-0_125-1},
	pages = {1},
}

@article{Schekochihin_Cowley-2006,
	title = {Turbulence, magnetic fields, and plasma physics in clusters of galaxies},
	volume = {13},
	issn = {1070-664X},
	url = {https://doi.org/10.1063/1.2179053},
	doi = {10.1063/1.2179053},
	urldate = {2026-03-25},
	journal = {Phys. Plasmas},
	author = {Schekochihin, A. A. and Cowley, S. C.},
	month = may,
	year = {2006},
	pages = {056501},
}

@article{Southwood_Kivelson-1993,
	title = {Mirror instability: 1. {Physical} mechanism of linear instability},
	volume = {98},
	copyright = {Copyright 1993 by the American Geophysical Union.},
	issn = {2156-2202},
	shorttitle = {Mirror instability},
	url = {https://onlinelibrary.wiley.com/doi/abs/10.1029/92JA02837},
	doi = {10.1029/92JA02837},
	urldate = {2024-11-27},
	journal = {J. Geophys. Res. Space Phys.},
	author = {Southwood, D. J. and Kivelson, M. G.},
	year = {1993},
	pages = {9181},
}

@article{Hasegawa-1969,
	title = {Drift {mirror} {instability} in the {magnetosphere}},
	volume = {12},
	issn = {0031-9171},
	url = {https://doi.org/10.1063/1.1692407},
	doi = {10.1063/1.1692407},
	urldate = {2025-04-14},
	journal = {Phys. Fluids},
	author = {Hasegawa, A.},
	month = dec,
	year = {1969},
	pages = {2642},
}

@article{Bott_etal-2024,
	title = {Kinetic stability of {Chapman}–{Enskog} plasmas},
	volume = {90},
	issn = {0022-3778, 1469-7807},
	url = {https://www.cambridge.org/core/product/identifier/S0022377824000308/type/journal_article},
	doi = {10.1017/S0022377824000308},
	langid = {english},
	urldate = {2025-04-14},
	journal = {J. Plasma Phys.},
	author = {Bott, A. F. A. and Cowley, S.C. and Schekochihin, A.A.},
	month = apr,
	year = {2024},
	pages = {975900207},
}

@article{Chandrasekhar_etal-1958,
	title = {The {stability} of the {pinch}},
	volume = {245},
	issn = {0080-46301364-5021},
	url = {https://ui.adsabs.harvard.edu/abs/1958RSPSA.245..435C},
	doi = {10.1098/rspa.1958.0094},
	urldate = {2025-11-06},
	journal = {Proc. R. Soc. Lond. A},
	author = {Chandrasekhar, S. and Kaufman, A. N. and Watson, K. M.},
	month = jul,
	year = {1958},
	pages = {435},
}

@article{Hellinger_etal-2006,
	title = {Solar wind proton temperature anisotropy: {Linear} theory and {WIND}/{SWE} observations},
	volume = {33},
	issn = {1944-8007},
	shorttitle = {Solar wind proton temperature anisotropy},
	url = {https://onlinelibrary.wiley.com/doi/abs/10.1029/2006GL025925},
	langid = {english},
	urldate = {2025-12-05},
	journal = {Geophys. Res. Lett.},
	author = {Hellinger, P. and Trávníček, P. and Kasper, J. C. and Lazarus, A. J.},
	year = {2006},
}

@article{Bale_etal-2009,
	title = {Magnetic {fluctuation} {power} {near} {proton} {temperature} {anisotropy} {instability} {thresholds} in the {solar} {wind}},
	volume = {103},
	doi = {10.1103/PhysRevLett.103.211101},
	urldate = {2025-12-05},
	journal = {Phys. Rev. Lett.},
	author = {Bale, S. D. and Kasper, J. C. and Howes, G. G. and Quataert, E. and Salem, C. and Sundkvist, D.},
	month = nov,
	year = {2009},
	pages = {211101},
}

@article{Chew_etal-1997,
	title = { The {Boltzmann} equation and the one-fluid hydromagnetic equations in the absence of particle collisions},
	volume = {236},
	url = {https://royalsocietypublishing.org/doi/10.1098/rspa.1956.0116},
	doi = {10.1098/rspa.1956.0116},
	urldate = {2025-07-12},
	journal = {Proc. R. Soc. Lond. A},
	author = {Chew, G. F. and Goldberger, M. L. and Low, F. E.},
	month = jan,
	year = {1956},
	pages = {112},
}

@article{Newcomb-1961,
	title = {Lagrangian and {Hamiltonian} methods in magnetohydrodynamics},
	url = {https://www.osti.gov/biblio/5034470},
	langid = {English},
	urldate = {2026-01-22},
    journal = {Nucl. Fusion Suppl., Part 2.},
	author = {Newcomb, W. A.},
	month = jun,
	year = {1961},
    pages = {451},
}

@incollection{Kulsrud-1964,
  author    = {Kulsrud, R.},
  title     = {General stability theory in plasma physics},
  booktitle = {Teoria dei plasmi, Rendiconti della scuola internazionale di fisica ``Enrico Fermi", XXV Corso},
  pages     = {54},
  year      = {1964},
  publisher = {Academic Press},
  address   = {New York}
}

@article{Sironi_Narayan-2015,
	title = {Electron Heating by the Ion Cyclotron Instability in Collisionless Accretion Flows. {I}. {Compression}-Driven Instabilities and the Electron Heating Mechanism},
	volume = {800},
	issn = {0004-637X},
	url = {https://dx.doi.org/10.1088/0004-637X/800/2/88},
	doi = {10.1088/0004-637X/800/2/88},
	langid = {english},
	urldate = {2024-11-21},
	journal = {Astrophys. J.},
	author = {Sironi, L. and Narayan, R.},
	month = feb,
	year = {2015},
	pages = {88},
}

@article{Gardner-1959,
	title = {Adiabatic {invariants} of {periodic} {classical} {systems}},
	volume = {115},
	copyright = {http://link.aps.org/licenses/aps-default-license},
	issn = {0031-899X},
	url = {https://link.aps.org/doi/10.1103/PhysRev.115.791},
	doi = {10.1103/PhysRev.115.791},
	langid = {english},
	urldate = {2025-10-30},
	journal = {Phys. Rev.},
	author = {Gardner, C. S.},
	month = aug,
	year = {1959},
	pages = {791},
}

@article{Kruskal-1962,
	title = {Asymptotic {theory} of {Hamiltonian} and other {systems} with all {solutions} {nearly} {periodic}},
	volume = {3},
	issn = {0022-2488, 1089-7658},
	url = {https://pubs.aip.org/jmp/article/3/4/806/227912/Asymptotic-Theory-of-Hamiltonian-and-other-Systems},
	doi = {10.1063/1.1724285},
	langid = {english},
	urldate = {2026-02-12},
	journal = {J. Math. Phys.},
	author = {Kruskal, M.},
	month = jul,
	year = {1962},
	pages = {806},
}

@article{Kunz_etal-2014,
	title = {Firehose and {mirror} {instabilities} in a {collisionless} {shearing} {plasma}},
	volume = {112},
	url = {https://link.aps.org/doi/10.1103/PhysRevLett.112.205003},
	doi = {10.1103/PhysRevLett.112.205003},
	urldate = {2024-11-21},
	journal = {Phys. Rev. Lett.},
	author = {Kunz, M. W. and Schekochihin, A. A. and Stone, J. M.},
	month = may,
	year = {2014},
	pages = {205003}, 
}

@article{Northrop_1963,
	title = {Adiabatic charged-particle motion},
	volume = {1},
	issn = {1944-9208},
	doi = {10.1029/RG001i003p00283},
	langid = {english},
	urldate = {2026-02-12},
	journal = {Rev. Geophys.},
	author = {Northrop, T. G.},
	year = {1963},
	pages = {283},
}

@article{Hellinger_Travnicek-2008,
	title = {Oblique proton fire hose instability in the expanding solar wind: {hybrid} simulations},
	volume = {113},
	issn = {2156-2202},
	shorttitle = {Oblique proton fire hose instability in the expanding solar wind},
	url = {https://onlinelibrary.wiley.com/doi/abs/10.1029/2008JA013416},
	langid = {english},
	urldate = {2025-10-30},
	journal = {J. Geophys. Res. Space Phys.},
	author = {Hellinger, P. and Trávníček, P. M.},
	year = {2008},
}

@article{Melville_etal-2016,
	title = {Pressure-anisotropy-driven microturbulence and magnetic-field evolution in shearing, collisionless plasma},
	volume = {459},
	issn = {0035-8711},
	url = {https://doi.org/10.1093/Mon. Not. R. Astron. Soc./stw793},
	doi = {10.1093/Mon. Not. R. Astron. Soc./stw793},
	urldate = {2024-11-21},
	journal = {Mon. Not. R. Astron. Soc.},
	author = {Melville, S. and Schekochihin, A. A. and Kunz, M. W.},
	month = jul,
	year = {2016},
	pages = {2701},
}

@article{Philippov_Kramer-2022,
	title = {Pulsar {magnetospheres} and {their} {radiation}},
	volume = {60},
	issn = {0066-4146},
	url = {https://ui.adsabs.harvard.edu/abs/2022ARA&A..60..495P},
	doi = {10.1146/annurev-astro-052920-112338},
	urldate = {2025-11-07},
	journal = {Annu. Rev. Astron. Astrophys.},
	author = {Philippov, A. and Kramer, M.},
	month = aug,
	year = {2022},
	pages = {495},
}

@article{Cerutti_Beloborodov-2017,
	title = {Electrodynamics of {pulsar} {magnetospheres}},
	volume = {207},
	issn = {1572-9672},
	url = {https://doi.org/10.1007/s11214-016-0315-7},
	doi = {10.1007/s11214-016-0315-7},
	langid = {english},
	urldate = {2026-02-12},
	journal = {Space Sci. Rev.},
	author = {Cerutti, B. and Beloborodov, A. M.},
	year = {2017},
	pages = {111},
}

@article{Rees_etal-1982,
	title = {Ion-supported tori and the origin of radio jets},
	volume = {295},
	issn = {1476-4687},
	doi = {10.1038/295017a0},
	langid = {english},
	urldate = {2026-02-12},
	journal = {Nature},
	author = {Rees, M. J. and Begelman, M. C. and Blandford, R. D. and Phinney, E. S.},
	year = {1982},
	pages = {17},
}

@article{Yuan_Narayan-2014,
	title = {Hot {accretion} {flows} {around} {black} {holes}},
	volume = {52},
	issn = {0066-4146, 1545-4282},
	url = {https://www.annualreviews.org/doi/10.1146/annurev-astro-082812-141003},
	doi = {10.1146/annurev-astro-082812-141003},
	langid = {english},
	urldate = {2026-02-12},
	journal = {Annu. Rev. Astron. Astrophys.},
	author = {Yuan, F. and Narayan, R.},
	year = {2014},
	pages = {529},
}

@article{Narayan_Yi-1995,
	title = {Advection-dominated {accretion}: {Underfed} {black} {holes} and {neutron} {stars}},
	volume = {452},
	issn = {0004-637X},
	url = {https://ui.adsabs.harvard.edu/abs/1995ApJ...452..710N},
	doi = {10.1086/176343},
	urldate = {2026-02-12},
	journal = {Astrophys. J.},
	publisher = {IOP},
	author = {Narayan, R. and Yi, I.},
	month = oct,
	year = {1995},
	pages = {710},
}

@article{Begelman_etal-1984,
	title = {Theory of extragalactic radio sources},
	volume = {56},
	url = {https://link.aps.org/doi/10.1103/RevModPhys.56.255},
	doi = {10.1103/RevModPhys.56.255},
	urldate = {2026-02-12},
	journal = {Rev. Mod. Phys.},
	publisher = {American Physical Society},
	author = {Begelman, M. C. and Blandford, R. D. and Rees, M.J.},
	month = apr,
	year = {1984},
	pages = {255},
}

@article{Hellinger-2007,
	title = {Comment on the linear mirror instability near the threshold},
	volume = {14},
	issn = {1070-664X},
	url = {https://doi.org/10.1063/1.2768318},
	doi = {10.1063/1.2768318},
	urldate = {2024-12-16},
	journal = {Phys. Plasmas},
	author = {Hellinger, P.},
	month = aug,
	year = {2007},
	pages = {082105},
}

@article{Yoon_etal-1993,
	title = {Effect of finite ion gyroradius on the fire‐hose instability in a high beta plasma},
	volume = {5},
	issn = {0899-8221},
	url = {https://doi.org/10.1063/1.860785},
	doi = {10.1063/1.860785},
	journal = {Phys. Fluids B},
	author = {Yoon, P. H. and Wu, C. S. and de Assis, A. S.},
	month = jul,
	year = {1993},
	pages = {1971},
}

@article{Schekochihin_etal-2010,
	title = {Magnetofluid dynamics of magnetized cosmic plasma: firehose and gyrothermal instabilities},
	volume = {405},
	issn = {0035-8711},
	url = {https://doi.org/10.1111/j.1365-2966.2010.16493.x},
	doi = {10.1111/j.1365-2966.2010.16493.x},
	journal = {Mon. Not. R. Astron. Soc.},
	author = {Schekochihin, A. A. and Cowley, S. C. and Rincon, F. and Rosin, M. S.},
	month = jun,
	year = {2010},
	pages = {291},
}

@article{Oieroset_etal-2001,
	title = {In situ detection of collisionless reconnection in the {Earth}'s magnetotail},
	volume = {412},
	copyright = {2001 Macmillan Magazines Ltd.},
	issn = {1476-4687},
	url = {https://www.nature.com/articles/35086520},
	doi = {10.1038/35086520},
	langid = {english},
	urldate = {2026-03-18},
	journal = {Nature},
	author = {Øieroset, M. and Phan, T. D. and Fujimoto, M. and Lin, R. P. and Lepping, R. P.},
	month = jul,
	year = {2001},
	pages = {414},
}

@article{Burch_etal-2016,
	title = {Magnetospheric {multiscale} {overview} and {science} {objectives}},
	volume = {199},
	issn = {1572-9672},
	url = {https://doi.org/10.1007/s11214-015-0164-9},
	doi = {10.1007/s11214-015-0164-9},
	langid = {english},
	urldate = {2026-03-18},
	journal = {Space Sci. Rev.},
	author = {Burch, J. L. and Moore, T. E. and Torbert, R. B. and Giles, B. L.},
	month = mar,
	year = {2016},
	pages = {5},
}

@article{Eckart-1940,
	title = {The {thermodynamics} of {irreversible} {processes}. {III}. {Relativistic} {theory} of the {simple} {fluid}},
	volume = {58},
	url = {https://link.aps.org/doi/10.1103/PhysRev.58.919},
	doi = {10.1103/PhysRev.58.919},
	urldate = {2025-07-29},
	journal = {Phys. Rev.},
	author = {Eckart, C.},
	year = {1940},
	pages = {919},
}

@article{Gedalin_Oiberman-1995,
	title = {Generally covariant relativistic anisotropic magnetohydrodynamics},
	volume = {51},
	url = {https://link.aps.org/doi/10.1103/PhysRevE.51.4901},
	doi = {10.1103/PhysRevE.51.4901},
	urldate = {2025-03-06},
	journal = {Phys. Rev. E},
	author = {Gedalin, M. and Oiberman, I.},
	month = may,
	year = {1995},
	pages = {4901}
}

@article{Galishnikova_etal-2023,
	title = {Polarized {anisotropic} {synchrotron} {emission} and {absorption} and {its} {application} to {black} {hole} {imaging}},
	volume = {957},
	issn = {0004-637X},
	url = {https://dx.doi.org/10.3847/1538-4357/acfa77},
	doi = {10.3847/1538-4357/acfa77},
	langid = {english},
	urldate = {2025-01-12},
	journal = {Astrophys. J.},
	author = {Galishnikova, A. and Philippov, A. and Quataert, E.},
	month = nov,
	year = {2023},
	pages = {103},
}

@article{Kargaltsev_etal-2015,
	title = {Pulsar-{wind} {nebulae}},
	volume = {191},
	issn = {1572-9672},
	url = {https://doi.org/10.1007/s11214-015-0171-x},
	doi = {10.1007/s11214-015-0171-x},
	langid = {english},
	urldate = {2025-10-30},
	journal = {Space Sci. Rev.},
	author = {Kargaltsev, O. and Cerutti, B. and Lyubarsky, Y. and Striani, E.},
	month = oct,
	year = {2015},
	pages = {391},
}

@article{Northrop_Teller-1960,
	title = {Stability of the {adiabatic} {motion} of {charged} {particles} in the {Earth}'s {field}},
	volume = {117},
	copyright = {http://link.aps.org/licenses/aps-default-license},
	issn = {0031-899X},
	url = {https://link.aps.org/doi/10.1103/PhysRev.117.215},
	doi = {10.1103/PhysRev.117.215},
	langid = {english},
	urldate = {2025-12-06},
	journal = {Phys. Rev.},
	author = {Northrop, T. G. and Teller, E.},
	month = jan,
	year = {1960},
	pages = {215},
}

@article{Vandervoort-1960,
	title = {The relativistic motion of a charged particle in an inhomogeneous electromagnetic field},
	volume = {10},
	issn = {0003-4916},
	url = {https://www.sciencedirect.com/science/article/pii/000349166090004X},
	doi = {10.1016/0003-4916(60)90004-X},
	urldate = {2025-11-24},
	journal = {Ann. Phys. (N. Y.)},
	author = {Vandervoort, P. O.},
	month = jul,
	year = {1960},
	pages = {401},
}

@article{Quataert-2003,
	title = {Radiatively {inefficient} {accretion} {flow} {models} of {Sgr} {A}*},
	volume = {324},
	issn = {1521-3994},
	url = {https://onlinelibrary.wiley.com/doi/abs/10.1002/asna.200385043},
	doi = {10.1002/asna.200385043},
	langid = {english},
	urldate = {2025-12-23},
	journal = {Astron. Nachr.},
	author = {Quataert, E.},
	year = {2003},
	pages = {435},
}

@article{Blandford_etal-2019,
	title = {Relativistic {jets} from {active} {galactic} {nuclei}},
	volume = {57},
	issn = {0066-4146, 1545-4282},
	doi = {10.1146/annurev-astro-081817-051948},
	langid = {english},
	urldate = {2025-12-23},
	journal = {Annu. Rev. Astron. Astrophys.},
	author = {Blandford, R. and Meier, D. and Readhead, A.},
	month = aug,
	year = {2019},
	pages = {467},
}

@article{Simon_Axford-1967,
	title = {{Thermal} {instability} {resulting} {from} {synchrotron} {radiation}},
	volume = {150},
	langid = {english},
    pages = {105},
	journal = {Astrophys. J.},
	author = {Simon, M. and Axford, W.I.},
	year = {1967},
}

@article{Bilbao_etal-2024,
	title = {Ring Momentum Distributions as a General Feature of {Vlasov} Dynamics in the Synchrotron Dominated Regime},
	volume = {31},
	issn = {1070-664X},
	url = {https://doi.org/10.1063/5.0206813},
	doi = {10.1063/5.0206813},
	urldate = {2024-11-21},
	journal = {Phys. Plasmas},
	author = {Bilbao, P. J. and Ewart, R. J. and Assunçao, F. and Silva, T. and Silva, L. O.},
	year = {2024},
	pages = {052112},
}

@article{Bilbao_Silvia-2023,
	title = {Radiation {reaction} {cooling} as a {source} of {anisotropic} {momentum} {distributions} with {inverted} {populations}},
	volume = {130},
	url = {https://link.aps.org/doi/10.1103/PhysRevLett.130.165101},
	doi = {10.1103/PhysRevLett.130.165101},
	urldate = {2025-04-08},
	journal = {Phys. Rev. Lett.},
	author = {Bilbao, P. J. and Silva, L. O.},
	year = {2023},
	pages = {165101},
}

@article{Zhdankin_etal-2023,
	title = {Synchrotron {firehose} {instability}},
	volume = {944},
	issn = {0004-637X},
	url = {https://dx.doi.org/10.3847/1538-4357/acaf54},
	doi = {10.3847/1538-4357/acaf54},
	langid = {english},
	urldate = {2024-11-21},
	journal = {Astrophys. J.},
	author = {Zhdankin, V. and Kunz, M.W. and Uzdensky, D.A.},
	year = {2023},
	pages = {24},
}

@inproceedings{Fonseca_etal-2002,
	address = {Berlin, Heidelberg},
	title = {{OSIRIS}: {A} {three}-{dimensional}, {fully} {relativistic} {particle} in {cell} {code} for {modeling} {plasma} {based} {accelerators}},
	isbn = {978-3-540-47789-1},
	shorttitle = {{OSIRIS}},
	doi = {10.1007/3-540-47789-6_36},
	langid = {english},
	booktitle = {Computational {Science} — {ICCS} 2002},
	publisher = {Springer},
	author = {Fonseca, R. A. and Silva, L. O. and Tsung, F. S. and Decyk, V. K. and Lu, W. and Ren, C. and Mori, W. B. and Deng, S. and Lee, S. and Katsouleas, T. and Adam, J. C.},
	year = {2002},
	pages = {342},
}

@article{Inchingolo_etal-2018,
	title = {Fully {kinetic} {large}-scale {simulations} of the {collisionless} {magnetorotational} {instability}},
	volume = {859},
	issn = {0004-637X},
	url = {https://doi.org/10.3847/1538-4357/aac0f2},
	doi = {10.3847/1538-4357/aac0f2},
	langid = {english},
	urldate = {2025-12-23},
	journal = {Astrophys. J.},
	author = {Inchingolo, G. and Grismayer, T. and Loureiro, N. F. and Fonseca, R. A. and Silva, L. O.},
	month = jun,
	year = {2018},
	pages = {149},
}

@article{Uzdensky_etal-2010,
	title = {Fast {magnetic} {reconnection} in the {plasmoid}-{dominated} {regime}},
	volume = {105},
	copyright = {http://link.aps.org/licenses/aps-default-license},
	issn = {0031-9007, 1079-7114},
	url = {https://link.aps.org/doi/10.1103/PhysRevLett.105.235002},
	doi = {10.1103/PhysRevLett.105.235002},
	langid = {english},
	urldate = {2026-02-19},
	journal = {	Phys. Rev. Lett.},
	author = {Uzdensky, D. A. and Loureiro, N. F. and Schekochihin, A. A.},
	month = dec,
	year = {2010},
	pages = {235002},
}

@article{Loureiro_etal-2007,
	title = {Instability of current sheets and formation of plasmoid chains},
	volume = {14},
	issn = {1070-664X},
	url = {https://doi.org/10.1063/1.2783986},
	doi = {10.1063/1.2783986},
	urldate = {2026-02-19},
	journal = {Phys. Plasmas},
	author = {Loureiro, N. F. and Schekochihin, A. A. and Cowley, S. C.},
	month = oct,
	year = {2007},
	pages = {100703},
}

@article{Werner_etal-2015,
	title = {The extent of power-law energy spectra in collisionless relativistic magnetic reconnection in pair plasmas},
	volume = {816},
	issn = {2041-8205},
	url = {https://doi.org/10.3847/2041-8205/816/1/L8},
	doi = {10.3847/2041-8205/816/1/L8},
	langid = {english},
	urldate = {2026-02-19},
	journal = {Astrophys. J.},
	author = {Werner, G. R. and Uzdensky, D. A. and Cerutti, B. and Nalewajko, K. and Begelman, M. C.},
	month = dec,
	year = {2015},
	pages = {L8},
}

@article{Sironi_Spitkovsky-2014,
	title = {Relativistic reconnection: an efficient source of non-thermal particles},
	volume = {783},
	issn = {2041-8205},
	url = {https://doi.org/10.1088/2041-8205/783/1/L21},
	doi = {10.1088/2041-8205/783/1/L21},
	langid = {english},
	urldate = {2026-02-19},
	journal = {Astrophys. J.},
	author = {Sironi, L. and Spitkovsky, A.},
	month = feb,
	year = {2014},
	pages = {L21},
}

@article{Drake_etal-2006,
	title = {Electron acceleration from contracting magnetic islands during reconnection},
	volume = {443},
	issn = {1476-4687},
	url = {https://doi.org/10.1038/nature05116},
	doi = {10.1038/nature05116},
	journal = {Nature},
	author = {Drake, J. F. and Swisdak, M. and Che, H. and Shay, M. A.},
	month = oct,
	year = {2006},
	pages = {553},
}

@article{Hakobyan_etal-2021,
	title = {Secondary energization in compressing plasmoids during magnetic reconnection},
	volume = {912},
	issn = {0004-637X},
	url = {https://doi.org/10.3847/1538-4357/abedac},
	doi = {10.3847/1538-4357/abedac},
	journal = {Astrophys. J.},
	author = {Hakobyan, H. and Petropoulou, M. and Spitkovsky, A. and Sironi, L.},
	month = may,
	year = {2021},
	pages = {48},
}

@article{Petropoulou_Sironi-2018,
	title = {The steady growth of the high-energy spectral cut-off in relativistic magnetic reconnection},
	volume = {481},
	issn = {0035-8711},
	url = {https://doi.org/10.1093/mnras/sty2702},
	doi = {10.1093/mnras/sty2702},
	journal = {Mon. Not. R. Astron. Soc.},
	author = {Petropoulou, M. and Sironi, L.},
	month = dec,
	year = {2018},
	pages = {5687},
}

@article{Le_etal-2009,
	title = {Equations of state for collisionless guide-field reconnection},
	volume = {102},
	url = {https://doi.org/10.1103/PhysRevLett.102.085001},
	doi = {10.1103/PhysRevLett.102.085001},
	journal = {Phys. Rev. Lett.},
	author = {Le, A. and Egedal, J. and Daughton, W. and Fox, W. and Katz, N.},
	month = feb,
	year = {2009},
	pages = {085001},
}

@article{Li_etal-2018,
	title = {Large-scale compression acceleration during magnetic reconnection in a low-beta plasma},
	volume = {866},
	issn = {0004-637X},
	url = {https://doi.org/10.3847/1538-4357/aae07b},
	doi = {10.3847/1538-4357/aae07b},
	journal = {Astrophys. J.},
	author = {Li, X. and Guo, F. and Li, H. and Li, S.},
	month = oct,
	year = {2018},
	pages = {4},
}

@article{Montag_etal-2017,
	title = {Impact of compressibility and a guide field on {Fermi} acceleration during magnetic island coalescence},
	volume = {24},
	issn = {1070-664X},
	url = {https://doi.org/10.1063/1.4985302},
	doi = {10.1063/1.4985302},
	journal = {Phys. Plasmas},
	author = {Montag, P. and Egedal, J. and Lichko, E. and Wetherton, B.},
	month = jun,
	year = {2017},
	pages = {062906},
}

@book{Synge-1957,
  author    = {Synge, J. L.},
  title     = {The relativistic gas},
  publisher = {North-Holland Publishing Company and Interscience Publishers},
  address   = {Amsterdam and New York},
  year      = {1957}
}

@book{Landau_Lifshitz-1987,
  author    = {Landau, L. D. and Lifshitz, E. M.},
  title     = {Fluid Mechanics},
  series    = {Course of Theoretical Physics},
  volume    = {6},
  year      = {1987},
  publisher = {Pergamon Press},
  address   = {Oxford}
}

@phdthesis{Simeon-2014,
	address = {United States -- California},
	type = {Ph.{D}.},
	title = {On the {acceleration}, {dynamics}, and {propagation} of {cosmic} {rays}},
	isbn = {979-8-6985-3771-7},
	langid = {English},
	urldate = {2026-02-25},
	school = {Stanford University},
	author = {Simeon, P.},
	year = {2014},
}

@article{Bott_etal-2025,
	title = {Thermodynamics and collisionality in firehose-susceptible high-$\beta$ plasmas},
	volume = {91},
	issn = {0022-3778, 1469-7807},
	doi = {10.1017/S0022377825100731},
	langid = {english},
	urldate = {2025-11-07},
	journal = {J. Plasma Phys.},
	author = {Bott, A. F. A. and Kunz, M. W. and Quataert, E. and Squire, J. and Arzamasskiy, L.},
	month = oct,
	year = {2025},
	pages = {E136},
}

@article{Wierzchucka_etal-2026,
  title   = {Relativistic Kinetic Magnetohydrodynamics},
  author  = {Wierzchucka, A. and Bilbao, P. J. and Uzdensky, D. A. and Schekochihin, A. A.},
  journal = {in preparation},
  year    = {2026}
}

@misc{Wierzchucka_etal-2026b,
    title   = {Two-Phase Structure of Synchrotron-Cooling-Unstable Relativistic Plasma},
	journal = {arXiv.org},
    author  = {Wierzchucka, A. and Bilbao, P. J. and Ewart, R. J. and Uzdensky, D. A. and Schekochihin, A. A.},
	year = {2026},
    eprint = {2609.07793},
    archivePrefix = {arXiv},
}

@article{Furth_etal-1963,
	title = {Finite‐{resistivity} {instabilities} of a {sheet} {pinch}},
	volume = {6},
	issn = {0031-9171},
	url = {https://doi.org/10.1063/1.1706761},
	doi = {10.1063/1.1706761},
	urldate = {2026-03-26},
	journal = {	Phys. Fluids},
	author = {Furth, H. P. and Killeen, J. and Rosenbluth, M. N.},
	month = apr,
	year = {1963},
	pages = {459},
}

@article{Hare_etal-2017,
	title = {Anomalous {heating} and {plasmoid} {formation} in a {driven} {magnetic} {reconnection} {experiment}},
	volume = {118},
	url = {https://link.aps.org/doi/10.1103/PhysRevLett.118.085001},
	doi = {10.1103/PhysRevLett.118.085001},
	urldate = {2026-03-28},
	journal = {Phys. Rev. Lett.},
	author = {Hare, J. D. and Suttle, L. and Lebedev, S. V. and Loureiro, N. F. and Ciardi, A. and Burdiak, G. C. and Chittenden, J. P. and Clayson, T. and Garcia, C. and Niasse, N. and Robinson, T. and Smith, R. A. and Stuart, N. and Suzuki-Vidal, F. and Swadling, G. F. and Ma, J. and Wu, J. and Yang, Q.},
	month = feb,
	year = {2017},
	pages = {085001},
}

@article{Totorica_etal-2017,
	title = {Particle acceleration in laser-driven magnetic reconnection},
	volume = {24},
	issn = {1070-664X, 1089-7674},
	doi = {10.1063/1.4978627},
	urldate = {2026-03-28},
	journal = {Phys. Plasmas},
	author = {Totorica, S. R. and Abel, T. and Fiuza, F.},
	month = apr,
	year = {2017},
	pages = {041408},
}

@article{Retino_etal-2007,
	title = {In situ evidence of magnetic reconnection in turbulent plasma},
	volume = {3},
	copyright = {2007 Springer Nature Limited},
	issn = {1745-2481},
	url = {https://www.nature.com/articles/nphys574},
	doi = {10.1038/nphys574},
	urldate = {2026-03-28},
	journal = {	Nat. Phys.},
	author = {Retinò, A. and Sundkvist, D. and Vaivads, A. and Mozer, F. and André, M. and Owen, C. J.},
	month = apr,
	year = {2007},
	pages = {235},
}

@misc{Ley_etal-2026,
	title = {On the double-adiabatic equations in the relativistic regime},
	journal = {arXiv.org},
	author = {Ley, F. and Tran, A. and Zweibel, E. G.},
	month = mar,
	year = {2026},
    eprint = {2603.25594},
    archivePrefix = {arXiv},
}

@article{Beskin_Kuznetsova-2000,
	title = {Grad-{Shafranov} {equation} with {anisotropic pressure}},
	volume = {541},
	issn = {0004-637X},
	doi = {10.1086/309438},
	urldate = {2026-03-28},
	journal = {Astrophys. J.},
	publisher = {IOP Publishing},
	author = {Beskin, V. S. and Kuznetsova, I. V.},
	month = sep,
	year = {2000},
	pages = {257},
}

@misc{Hegade_Stone-2026,
	title = {Kinetic magnetohydrodynamics and {Landau} fluid closure in relativity},
	journal = {arXiv.org},
	author = {Hegade, K. R. and Stone, J. M.},
	month = apr,
	year = {2026},
    eprint = {2604.05058},
    archivePrefix = {arXiv},
}

@article{Bret_Narayan-2018,
	title = {Density jump as a function of magnetic field strength for parallel collisionless shocks in pair plasmas},
	volume = {84},
	copyright = {https://www.cambridge.org/core/terms},
	issn = {0022-3778, 1469-7807},
	url = {https://www.cambridge.org/core/product/identifier/S0022377818001125/type/journal_article},
	doi = {10.1017/S0022377818001125},
	language = {en},
	urldate = {2026-07-15},
	journal = {J. Plasma Phys.},
	author = {Bret, A. and Narayan, R.},
	month = dec,
	year = {2018},
	pages = {905840604},
}

@article{Sironi_Spitkovsky-2009,
	title = {Particle acceleration in relativistic magnetized collisionless pair shocks: Dependence of shock acceleration on magnetic obliquity},
	volume = {698},
	doi = {10.1088/0004-637X/698/2/1523},
	journal = {Astrophys. J.},
	author = {Sironi, L. and Spitkovsky, A.},
	month = jun,
	year = {2009},
	pages = {1523},
}

@article{Sironi_etal-2013,
	title = {The maximum energy of accelerated particles in relativistic collisionless shocks},
	volume = {771},
	doi = {10.1088/0004-637X/771/1/54},
	journal = {Astrophys. J.},
	author = {Sironi, L. and Spitkovsky, A. and Arons, J.},
	month = jun,
	year = {2013},
	pages = {54},
}

\end{document}